\documentclass[%
 reprint,
 amsmath,amssymb,
 aps,
]{revtex4-2}

\usepackage[utf8]{inputenc}

\usepackage{graphicx}% Include figure files
\usepackage{dcolumn}% Align table columns on decimal point
\usepackage{bm}% bold math
\usepackage{siunitx}
\usepackage{chemformula}
\usepackage{txfonts}
\usepackage{amsmath}
\usepackage{braket}
\usepackage{float}

\usepackage{hyperref}
\renewcommand{\url}[1]{}
\begin{document}

\preprint{APS/123-QED}

%\linenumbers

\title{Two-qubit logic and teleportation with mobile spin qubits in silicon}% Force line breaks with \\
% other candidates here
% "Two-qubit gates and teleportation between sparse spin qubits in silicon"
% "Shuttling-based two-qubit logic and teleportation with distant spins in silicon"

% High-fidelity bucket-brigade and conveyor-belt spin shuttling in silicon
\author{Y. Matsumoto$^{1 \dag}$}
\author{M. De Smet$^{1 \dag}$}
\author{L. Tryputen$^{2}$}
\author{S.L. de Snoo$^{1}$}
\author{S.V. Amitonov$^{2}$}
\author{A. Sammak$^{2}$}
\author{M. Rimbach-Russ$^{1}$}
\author{G. Scappucci$^{1}$}
\author{L.M.K. Vandersypen$^{1}$}
\email{L.M.K.Vandersypen@tudelft.nl}

%  \email{Second.Author@institution.edu}
\affiliation{$^{1}$QuTech and Kavli Institute of Nanoscience, Delft University of Technology, Lorentzweg 1, 2628 CJ Delft, The Netherlands \\ $^{2}$QuTech and Netherlands Organization for Applied Scientific Research (TNO), Delft, The Netherlands} % This line break forced with \textbackslash\textbackslash

\affiliation{$^{\dag}$ These authors contributed equally}
% \def\thefootnote{*}\footnotetext{These authors contributed equally to this work}

% \collaboration{MUSO Collaboration}%\noaffiliation

%  \homepage{http://www.Second.institution.edu/~Charlie.Author}
% \affiliation{
%  Second institution and/or address\\
%  This line break forced% with \\
% }%
% \affiliation{
%  Third institution, the second for Charlie Author
% \affiliation{%
%  Authors' institution and/or address\\
%  This line break forced with \textbackslash\textbackslash
% }%

% \collaboration{CLEO Collaboration}%\noaffiliation

\date{\today}% It is always \today, today,

\begin{abstract}

The scalability and power of quantum computing architectures depend critically on high-fidelity operations and robust and flexible qubit connectivity. In this respect, mobile qubits are particularly attractive as they enable dynamic and reconfigurable qubit arrays. This approach allows quantum processors to adapt their connectivity patterns during operation, implement different quantum error correction codes on the same hardware, and optimize resource utilization through dedicated functional zones for specific operations like measurement or entanglement generation. Such flexibility also relieves architectural constraints, as recently demonstrated in atomic systems based on trapped ions and neutral atoms manipulated with optical tweezers. In solid-state platforms, highly coherent shuttling of electron spins was recently reported. A key outstanding question is whether it may be possible to perform quantum gates directly on the mobile spins.
In this work, we demonstrate two-qubit operations between two electron spins carried towards each other in separate traveling potential minima in a semiconductor device. We find that the interaction strength is highly tunable by their spatial separation, achieving an average two-qubit gate fidelity of about 99\%.
Additionally, we implement conditional post-selected quantum state teleportation between spatially separated qubits with an average gate fidelity of 87\%, showcasing the potential of mobile spin qubits for non-local quantum information processing. We expect that operations on mobile qubits will become a universal feature of future large-scale semiconductor quantum processors.

\end{abstract}

\maketitle

%\tableofcontents
% \protect\\ The line
% break was forced \lowercase{via} \textbackslash\textbackslash

\section{\label{sec:Introduction}Introduction}
Quantum computing offers the promise to solve complex problems that are intractable for classical computers. As quantum processors scale up, maintaining high connectivity between qubits becomes crucial for implementing effective error correction schemes~\cite{bravyi_high-threshold_2024, xu_constant-overhead_2024,goto_manyhypercube_2024}. However, traditional architectures are often restricted to interactions between nearest neighbors, constraining the options for quantum error correction codes and potentially increasing overhead. Mobile qubits offer a promising alternative by enabling flexible connectivity between qubits, thus reducing the overhead associated with error correction schemes~\cite{pino_demonstration_2021,sterk_closed_loop_2022,bluvstein_quantum_2022, bluvstein_logical_2023}.

Among the various quantum computing platforms, gate-defined semiconductor spin qubits \cite{vandersypen_quantum_2019} have emerged as a promising candidate. These qubits offer a compelling combination of extended coherence times \cite{veldhorst_addressable_2014}, high-fidelity operations \cite{yoneda_quantum-dot_2018, yang_silicon_2019, lawrie_simultaneous_2023, xue_quantum_2022, noiri_fast_2022, mills_two-qubit_2022,wang_operating_2024,tanttu_assessment_2024,steinacker_violating_2024},  compatibility with established semiconductor manufacturing techniques \cite{maurand_cmos_2016, zwerver_qubits_2022,george_12spin_2024,steinacker_300mm_2024,huckemann_industrially_2024} and the potential for high-temperature operation \cite{undseth_hotter_2023, huang_high-fidelity_2024}.

Inspired by mobile qubit approaches in atomic systems, where trapped ions~\cite{pino_demonstration_2021,sterk_closed_loop_2022} and neutral atoms~\cite{bluvstein_quantum_2022, bluvstein_logical_2023} have demonstrated the power of reconfigurable qubit arrays, the question arises whether mobile semiconductor spin qubits offer a route to realize flexible connectivity in a solid-state platform. In recent experiments, a traveling wave potential, generated by phase-shifted sinusoidal signals applied to successive gate electrodes, transports the spin qubit within a moving quantum dot \cite{taylor_fault-tolerant_2005,seidler_conveyor-mode_2022, struck_spin-epr-pair_2024, xue_sisige_2024}. With this so-called conveyor shuttling method a 99.5\% fidelity was achieved when shuttling across an effective 10 $\mu$m distance in under 200 ns \cite{desmet_highfidelityshuttle_2024}.

\begin{figure*}[t]
\includegraphics[width=\textwidth]{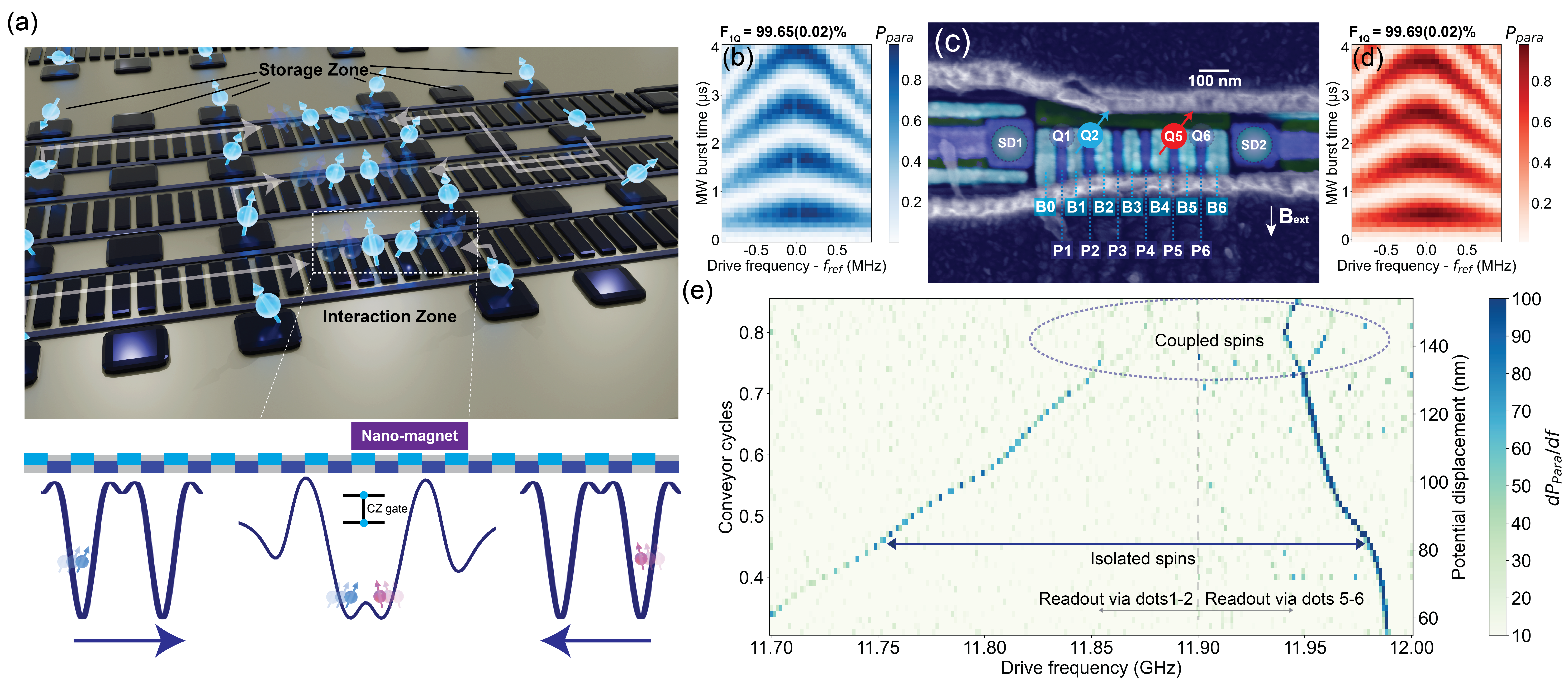}
\caption{\label{fig:fig1} \textbf{Mobile spin qubits and shuttling-based architecture.}
(a) Conceptual architecture for a scalable mobile spin qubit processor based on conveyor-mode shuttling. Qubits can be transported between static storage zones (static dots) and pairs of adjacent conveyor channels that meet at shared interaction zones. Two-qubit operations are performed by simultaneously shuttling two qubits inside the same channel to an interaction zone. This design enables efficient resource sharing. Vertical transport between parallel conveyor channels, passing through vacant storage zones, allows for flexible connectivity across the array (see white arrows). Readout zones with sensing dots and ancilla qubits at array endpoints enable parallel measurements of multiple qubits. 
b) d) Parallel spin probability plotted as a function of drive frequency detuning from $f_{ref}$ (11.657 GHz for Q2 and 11.999 GHz for Q5) and microwave burst time reveals Rabi chevron patterns from single-qubit rotations for Q2 and Q5. The average single-qubit gate fidelities, extracted by randomized benchmarking, are shown above the respective chevron patterns. 
c) False-colored scanning electron microscope (SEM) image of a nominally identical device to the one used in this work. The colors indicate different metallization layers. Six plungers (cyan), seven barriers (blue), and two screening gates (dark green) form two distant double dots 1-2 and 5-6 (indicated by numbered circles). Two sensing dots (SD) are placed at both ends of the array. A cobalt micromagnet, shown in (dark blue), is placed on top of the active area. 
e) EDSR spectroscopy as a function of microwave drive frequency and the number of conveyor cycles (left axis), with the corresponding nominal  displacement of the potential minima (right axis). The signal shows the numerical derivative of the parallel spin probability with respect to the microwave frequency, measured via dots 1-2 (11.70-11.90 GHz) and dots 5-6 (11.90-12.00 GHz). The dashed vertical line at 11.90 GHz indicates the boundary between the two measurement configurations.}
\end{figure*}

Building on these advances, we envision scalable mobile spin qubit architectures based on conveyor mode shuttling as shown in Figure \ref{fig:fig1}a, where qubits can be selectively transported between storage zones, and where interaction regions are formed by pairs of independent conveyor channels. This approach enables efficient resource sharing and flexible qubit connectivity through shared control lines and sparse storage zones. It may also facilitate  uniformly high gate fidelities by performing operations in optimized local electromagnetic environments. The fundamental building block and key challenge for realizing such architectures is the ability to precisely control the interaction between two mobile spin qubits by selectively bringing them together using independent conveyor channels and to achieve high gate fidelities\cite{kunne_cvarchitecture_2024}.

% ###
In this work, we focus on demonstrating this critical operation using a linear array implementation. We simultaneously shuttle two spin qubits in conveyor-mode towards each other, such that their wavefunctions begin to overlap and a two-qubit exchange interaction is activated. By systematically examining the exchange interaction over various shuttle distances and barrier voltage configurations, we seek to identify the conditions under which tunable and high-fidelity two-qubit gates can be realized. Moreover, we explore a regime in which merging extremely elongated quantum dot potentials may lead to exchange interaction saturation~\cite{wigner_2021,wigner_exp_2021,jang_wignerexp_2023}. Additionally, we employ the two-qubit gate on mobile spins to entangle and separate two spins, and use this entanglement to realize conditional post-selected quantum state teleportation of single-spin states, effectively spanning five quantum dots. 

\section{\label{sec:Results}Results}

\subsection{\label{subsec:mobile}Mobile spin qubits and shuttling-based architecture}

We operate mobile spin qubits in a six-quantum-dot array fabricated on an isotopically purified $^{28}$Si/SiGe heterostructure featuring a \SI{7}{\nano \meter} quantum well~\cite{degli_esposti_low_2024} (Fig.~\ref{fig:fig1}b). The quantum dots are defined by three layers of Ti:Pd gates separated by \ch{Al2O3}, consisting of screening, plunger (P), and barrier (B) gates. For readout, sensing dots are positioned at both ends of the array. A cobalt micromagnet integrated on top of the device provides the magnetic field gradient necessary for single-qubit control by means of electric-dipole spin resonance (EDSR)~\cite{obata_coherent_2010}. We operate the device in an in-plane magnetic field of \SI{260}{\milli T} at a temperature of \SI{200}{\milli K}~\cite{undseth_hotter_2023}.

To test two-qubit operations on mobile qubits, we will load spins Q2 and Q5 from dots 2 and 5 into two independent conveyor potentials traveling inside the channel in between. Spins in dots 1 and 6, Q1 and Q6, serve as stationary readout ancillas. Readout relies on a variant of Pauli spin blockade, which reveals the parity of Q1-Q2 and Q5-Q6. We observe dephasing times $T_{2}^{*}$ of 5.39$\pm$0.29 $\mu$s and 7.14$\pm$0.36 $\mu$s for Q2 and Q5 respectively (measured in the static dots), extending to $T_2^{echo}$ times of 18.15$\pm$1.06 $\mu$s and 39.80$\pm$2.76 $\mu$s under Hahn-echo sequences. The chevron patterns shown in Figs.~\ref{fig:fig1}b and \ref{fig:fig1}d demonstrate precise single-qubit control, with randomized benchmarking yielding single-qubit gate fidelities of 99.65$\pm$0.02\% and 99.69$\pm$0.02\% for Q2 and Q5, respectively. When operating the qubits simultaneously, the fidelities decrease to 99.03$\pm$0.17\% and 99.51$\pm$0.06\% for Q2 and Q5, resulting in an overall joint fidelity for simultaneous operation of 98.54$\pm$0.18\%.

For conveyor-mode shuttling, multiple phase-shifted sinusoidal signals are applied to plungers and barrier gates. As in~\cite{desmet_highfidelityshuttle_2024}, two tones are applied, with frequeny $f$ ($f/2$) and a spatial period of four (eight) gate electrodes. We adjust the DC gate voltages in the channel such that the background potential landscape is approximately flat. As a result, a traveling wave potential carries electrons within a moving potential well. This method allows for continuous control of the electron positions along the array. In this work, we operate two mobile qubits using two moving potentials, which move from the positions of dot 2 and dot 5 towards the center. This is achieved by applying sine waves with phase offsets that increase symmetrically from P2 and P5 toward the central barrier B3 (see supplementary Table~\ref{tab:conveyor_params}).

To characterize the simultaneous transport of two mobile spin qubits, we perform EDSR spectroscopy as a function of the number of conveyor cycles applied (Figure \ref{fig:fig1}e), where one conveyor cycle corresponds to a nominal displacement of 180 nm, which is twice the plunger gate pitch (we refer to the number of conveyor cycles of the primary frequency component $f$). In the left panel, Q2 is initialized in the spin-down state while Q5 is prepared in a random state; in the right panel, Q2 is prepared in a random state while Q5 is initialized in the spin-up state. For each configuration, we shuttle both spins using conveyor mode shuttling, apply a microwave burst for EDSR, return the spins to dots 2 and dot 5, and perform readout. The observed shift in resonance frequency with increasing conveyor cycles indicates a gradual spin displacement along the channel. The different slopes for Q2 and Q5 arise from the different magnetic field gradients experienced by the qubits as they are shuttled toward the center. As the electrons approach each other, the EDSR spectra reveal split lines, indicating that the two spins interact, such that the resonance frequency of one is dependent on the state of the other.

\begin{figure*}[t]
\includegraphics[width=1\textwidth]{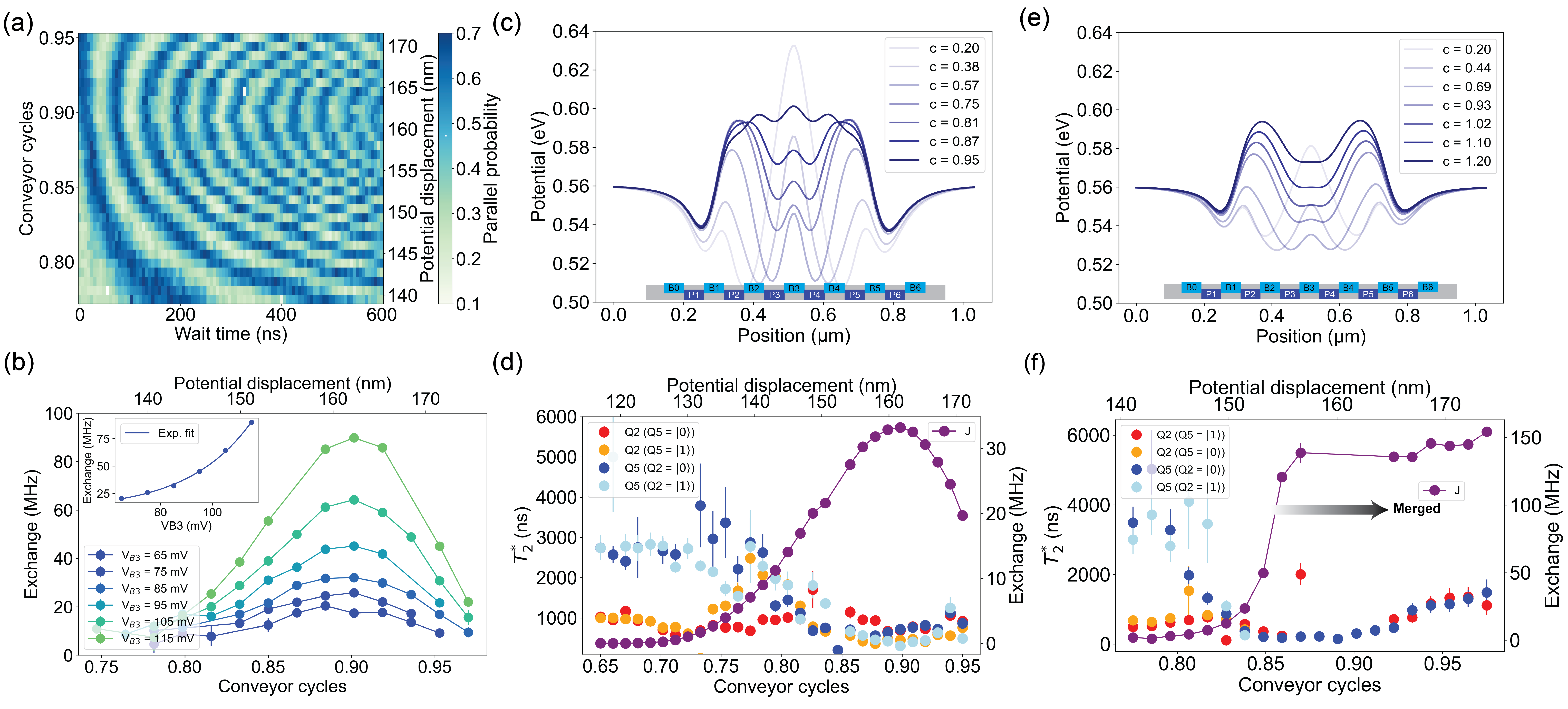}
\caption{\label{fig:fig2} \textbf{Exchange and coherence of mobile spin qubits}.
a) Two-dimensional map showing DCPhase oscillations as a function of wait time and conveyor cycles (left axis), with corresponding nominal potential displacement (right axis), with a 65 mV pulsed voltage offset on B3. See main text for further details.
b) Exchange coupling $J$ versus conveyor cycles (bottom axis), with corresponding nominal potential displacement (top axis), at different pulsed voltage offsets on $V_{\text{B3}}$, and extracted from DCPhase oscillations as in panel (a). The inset shows the peak exchange strength versus the voltage on B3, and an exponential fit.
c) Simulated potential profiles showing the evolution of the two moving potential minima during shuttling, where $c$ represents the number of conveyor cycles applied. A cycle is defined by one full period of the primary frequency component $f$.
d) Measured dephasing times $T_2^*$ for both qubits under different states of the other spin: Q2(Q5=$\ket{0}$), Q2(Q5=$\ket{1}$), Q5(Q2=$\ket{0}$), and Q5(Q2=$\ket{1}$), plotted against conveyor cycles (bottom axis), with corresponding nominal potential displacement (top axis). The right axis shows the exchange coupling $J$ extracted from the measured difference in Ramsey frequency depending on the state of the other spin. A 9.5 mV pulse offset is applied to B3, consistent with panel (c).
e) Simulated potential profiles at different conveyor cycles $c$ for the elongated dot configuration. As the confinement potential evolves with increasing $c$, the elongated potential minima merge into a single even more elongated potential, supporting strong Coulomb interactions between the electrons, which favors the formation of a Wigner molecule. 
f) Exchange coupling and dephasing times $T_{2}^{*}$ for Q2(Q5=$\ket{0}$), Q2(Q5=$\ket{1}$), Q5(Q2=$\ket{0}$), and Q5(Q2=$\ket{1}$) in the merged configuration of panel (e), showing exchange saturation at larger conveyor cycles. The exchange coupling is extracted from DCPhase oscillations in this condition.
}
\end{figure*}
\subsection{\label{subsec:EX}Exchange and coherence of mobile spin qubits}

Next, we evaluate the controllability of the exchange interaction achieved through shuttling. To characterize the exchange interaction, we employ a decoupled controlled-phase (DCPhase) sequence: after initialization, an $R_x (\pi/2)$ pulse is applied to Q2, followed by shuttling both qubits together to activate exchange coupling for a duration $t/2$ and shuttling the spins back to their starting position. $R_x (\pi)$ gates are then applied to both spins in their respective dots, followed by another exchange interaction period $t/2$. Finally, another $R_x (\pi/2)$ pulse is applied to Q2 before measuring both qubits using parity readout against their reference spins. Figure 2a presents the measured DCPhase oscillations versus conveyor cycles with a 65 mV pulse offset applied to barrier gate B3 (compared to the flatband voltage setting). The oscillation frequency, which directly reflects the exchange amplitude $J$, increases smoothly as the potential minima are moved towards each other before plateauing and subsequently decreasing.

Figure 2b shows the estimated exchange strength from DCPhase measurements performed at different B3 voltage offsets (ranging from 65 to 115 mV). For each curve, $J$ reaches a maximum and then slightly decreases. This is counterintuitive in the picture where two potential minima progressively move towards each other. It results from the fact that the central barrier gate B3 receives half the conveyor amplitude (and a negative pulse offset) compared to the other conveyor gates. This keeps the minima from the two conveyor channels separated by a tunnel barrier while still allowing a controlled exchange coupling. Figure 2c illustrates this through simulated potential profiles at different conveyor cycles $c$. As seen in the inset to Figure 2b, $J$  increases exponentially with the voltage offset on B3, consistent with the Fermi-Hubbard model describing tunnel-coupled quantum dots~\cite{meunier_cphase_2011}. At the start of the conveyor (zero conveyor cycles), $J$ is too small to measure. The exchange strength between the mobile qubits can be tuned up to 90 MHz through control of both the number of conveyor cycle c and the barrier voltage. While higher exchange strengths are achievable, they were difficult to measure due to their rapid decay under these conditions.

The two-qubit gate fidelity that can be achieved depends on the balance between $J$ and $T_{2}^{*}$.
Figure 2d presents the $T_{2}^{*}$ dephasing times of Q2 and Q5, along with the corresponding exchange strength $J$, as a function of the number of conveyor cycle. As $c$ increases, leading to a stronger exchange interaction and faster gate operation, we observe a decrease in $T_{2}^{*}$ for both qubits. The difference in $T_{2}^{*}$ depending on the state ($\ket{0} \text{ or } \ket{1}$) of the other qubit can be attributed to the spatial magnetic field gradient in the device~\cite{xue_quantum_2022}. We note that $T_{2}^{*}$ during shuttling in the device typically exceeds the static $T_{2}^{*}$ measured at fixed positions along the channel~\cite{desmet_highfidelityshuttle_2024}. This is due to motional averaging as the qubit samples different local environments at a rate faster than the correlation time of both the nuclear field fluctuations and charge noise~\cite{struck_spin-epr-pair_2024,mokeev_modeling_2023}. Based on this characterization, we identify an operating regime that balances a strong enough exchange coupling for fast gates while maintaining sufficient coherence times. This condition, corresponding to 0.9 conveyor cycles and an exchange coupling of 33 MHz, enables the high-fidelity two-qubit operations demonstrated in Figure 3.

Before proceeding to quantifying the gate fidelity, we investigate the exchange strength and spin coherence in a configuration where two elongated traveling potential minima are created by applying a single sinusoidal wave having a spatial period corresponding to 8 gate electrodes (once again with phase offsets that increase symmetrically from the outer gates toward the center; here B3 receives a pulse offset of -10 mV). When the two elongated potentials overlap significantly at the center, the system transitions away from a double-dot regime where the Fermi-Hubbard description is applicable. Numerical simulations (Figure 2e) reveal how the potential profile transitions from a barrier-controlled double-well configuration into a single highly elongated potential. 

Figure 2f shows that in this regime, $J$ initially exhibits the standard exponential increase expected for a double quantum dot (up to $\approx 0.86$ conveyor cycles), where the potential barrier dictates the coupling. Beyond this point, as the two elongated potentials merge, $J$ no longer increases exponentially but instead saturates. This saturation may originate from strong electron-electron interactions within the merged elongated dot~\cite{wigner_2021,wigner_exp_2021,jang_wignerexp_2023}. Notably, once in this merged regime, $J$ becomes relatively insensitive to barrier voltage variations, enabling an enhancement of $T_{2}^{*}$ while still maintaining substantial exchange coupling. Unfortunately, before the conveyor has reached this favorable condition, it passes through a region where $T_2^*$ is very short, resulting in overall low two-qubit gate fidelities. In fact, when measuring Q2's $T_2^*$ with Q5 initialized in $\ket{0}$, or Q5's $T_2^*$ with Q2 in $\ket{1}$, we cannot even accurately estimate $T_{2}^*$ because coherence is lost rapidly before $J$ reaches the saturation regime. 
In the region with very short $T_2^*$, the frequency shifts due to both the magnetic field gradient and the exchange interaction add constructively, as voltage fluctuations or charge noise simultaneously affect both the Zeeman energy splitting and the exchange interaction in a correlated way~\cite{xue_quantum_2022}. %Future implementations could benefit from operating in regions with reduced magnetic field gradients, allowing for robust high-fidelity two-qubit gates.

\begin{figure*}
\includegraphics[width=0.95\textwidth]{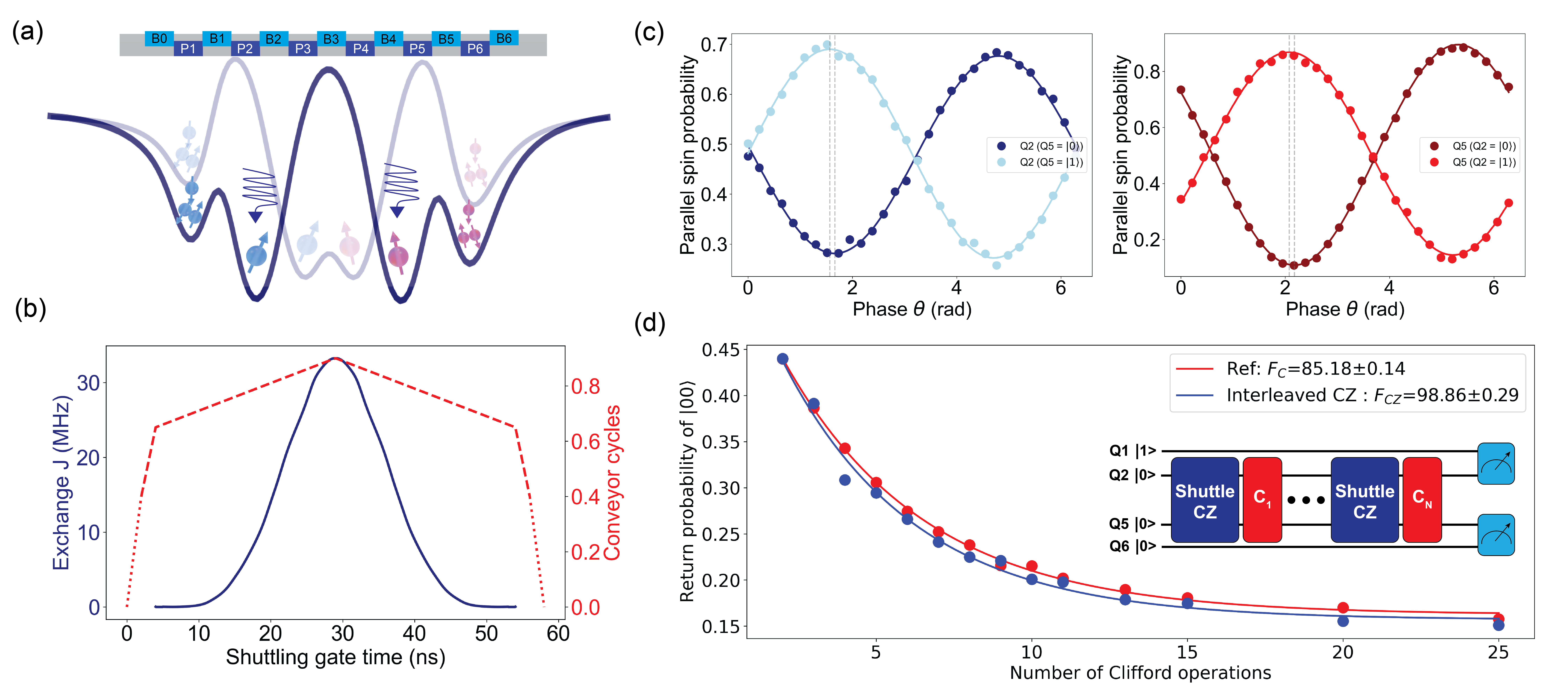}
\caption{\label{fig:fig3} \textbf{Fidelity benchmarking of shuttling-based CZ gate}
a) Schematic illustration of the gate electrodes and simulated potential profiles with static dots (dark line) and the conveyor during the interaction phase (faint line). 
b) Time evolution of the exchange coupling strength $J$ (blue, left axis) and the nominal electron displacement expressed in conveyor cycles (red dashed lines, right axis) throughout the CZ gate operation. The dotted lines indicate the loading and unloading phases between the initial positions of the conveyor’s potential minima and the static dots. The pulse shape is designed to achieve adiabatic control of the exchange interaction.
c) Calibration measurements of the CZ gate showing parallel spin probability oscillations for Q2 (left) and Q5 (right) as a function of an applied virtual phase shift $\theta$. The measurements are performed with the other qubit initialized in either $\ket{0}$ or $\ket{1}$ state, as indicated.
d) Results of interleaved randomized benchmarking, showing the return probability versus the number of Clifford operations for both reference (red) and interleaved CZ gate (blue) sequences. The schematic illustrates the IRB protocol where shuttling-based CZ gates (Shuttle CZ) are interleaved between random Clifford operations ($C_{1}$ to $C_{N}$) for the interleaved measurements, while reference measurements are performed with only the red Clifford operations.}
\end{figure*}

\subsection{\label{subsec:CV}Fidelity benchmarking of shuttling based CZ gate}
Returning to the conveyor configuration of Figure 2c,d, we evaluate the performance of a shuttling-based conditional-Z (CZ) gate. We set the target maximum exchange strength of 33 MHz such that the exchange strength is much smaller than the Zeeman energy difference between Q2 and Q5 after they have been brought into each other's vicinity. The Zeeman energy difference suppresses the flip-flop terms of the exchange interaction, leaving only the ZZ term active.

Figure 3a illustrates schematically how the device is operated. The four qubits (Q1, Q2, Q5, Q6) are initialized in the $\lvert$1000⟩ state, where Q1 and Q6 serve once again as ancilla qubits for parity readout. Microwave bursts for single-qubit control are applied while qubits Q2 and Q5 are confined in the static dots 2 and 5. For the two-qubit CZ gate, conveyor-mode shuttling carries Q2 and Q5 towards each other and back for a controlled duration. 

Figure 3b shows the target potential displacement over time and the resulting $J$ throughout the two-qubit gate (interpolated from the datapoints in Fig. 2d). The pulse sequence is designed to balance speed and adiabaticity, consisting of the following main stages: first, a 2ns loading phase (dotted lines) carries the electrons from static dot 2 (Q2) and 5 (Q5) to the initial conveyor potential at $c=0.4$ (where $c=0$ would correspond to a conveyor minimum centered below P2 and P5); second a fast 2 ns approach phase using a 125 MHz conveyor frequency minimizes dephasing during initial transport (from $c=0.4$ to $c=0.65$); third, a 25 ns interaction phase using a 10 MHz conveyor frequency where the inter-qubit distance is adiabatically reduced to activate the exchange interaction; and finally, a symmetric return sequence. Including the 2 ns loading and approach phases, the total CZ gate time is 58 ns. The two spins are initially 270 nm apart, spanning four quantum dots.

The CZ gate calibration uses a set of measurements similar to those shown in Figure 3c, exploring different center barrier voltages (see Supplementary Figure~\ref{supp:fig:RB}). In the measurement sequence, one spin (Q2 for the left panel, Q5 for the right panel) is prepared in a superposition state using an $R_x (\pi/2)$ gate, while the other spin is initialized in either $\ket{0}$ or $\ket{1}$. The spins are then shuttled to the center where the exchange interaction is activated, followed by return shuttling to their initial positions. Before measurement, a virtual $R_{Z}(\theta)$ rotation is applied through microwave phase adjustment, followed by another $R_x (\pi/2)$ gate. The left panel displays the parallel spin probability for Q2 as a function of $\theta$, with the dark blue curve representing Q5 in $\ket{0}$ and the light blue curve for Q5 in $\ket{1}$. The right panel shows the complementary data for Q5, with the dark red curve representing Q2 in $\ket{0}$ and the red curve Q2 in $\ket{1}$. The two cases are (nearly) out of phase, as expected for a properly calibrated CZ gate. These same measurements also serve to calibrate the single-qubit phase shifts accumulated during shuttling. This is done by varying $\theta$ such that the largest contrast between the light and dark datapoints is reached. Looking closely, in Fig. 3c, a slight deviation from a $\pi$ controlled phase is seen even though the gate is properly calibrated. This arises from additional phase accumulation due to the extra microwave burst~\cite{undseth_hotter_2023} that is applied only when preparing the other qubit in $\ket{1}$, rather than from CZ gate imperfections (see Supplementary Fig.~\ref{supp:fig:heating}).

Fig. 3d shows the results of interleaved randomized benchmarking (IRB) measurements. For every reference and interleaved measurement, we produce 120 distinct sequences at random. We performed 800 single shots for each sequence, with an average of approximately 250 single shots being post-selected~\cite{philips_universal_2022}. The data is plotted as the return probability to the initial state $\ket{00}$ versus the number of Clifford gates $N$, for both the reference sequence and the interleaved benchmarking sequence. The decay of the return probability is analyzed to extract the fidelity of the Clifford gates  $F_C$  and the fidelity of the CZ gate $F_{CZ}$.
The estimated fidelity $F_{CZ}$= 98.86 $\pm$ 0.29\% demonstrates the high performance of the shuttling-based CZ gate. The average Clifford gate fidelity from the reference measurement is 85.18\%, which is consistent with the expected fidelity based on the gate composition. Using the formula $F_C = 1 - (1.5r_{CZ} + 8.25r_{SQ})$, with an IRB-extracted CZ error rate $r_{CZ}$ = 0.0114 and a joint single-qubit error rate with simultaneous driving $r_{SQ}$ = 0.0146, we estimate $F_C$ = 85.53\%.

To understand the sources of infidelity in the CZ gate, we perform a theoretical analysis and numerical simulations. The calculations estimate infidelity contributions of only $\sim$ 0.01\% from coherent errors induced by the unwanted flip-flop terms and of $\sim$ 0.22\% from incoherent errors due to low-frequency noise during the CZ operation~\cite{wang_operating_2024} (see supplementary information). Additional error sources include heating effects from microwave bursts and high-frequency noise in the exchange control~\cite{undseth_hotter_2023}.

\begin{figure*}[t]
\includegraphics[width=1\textwidth]{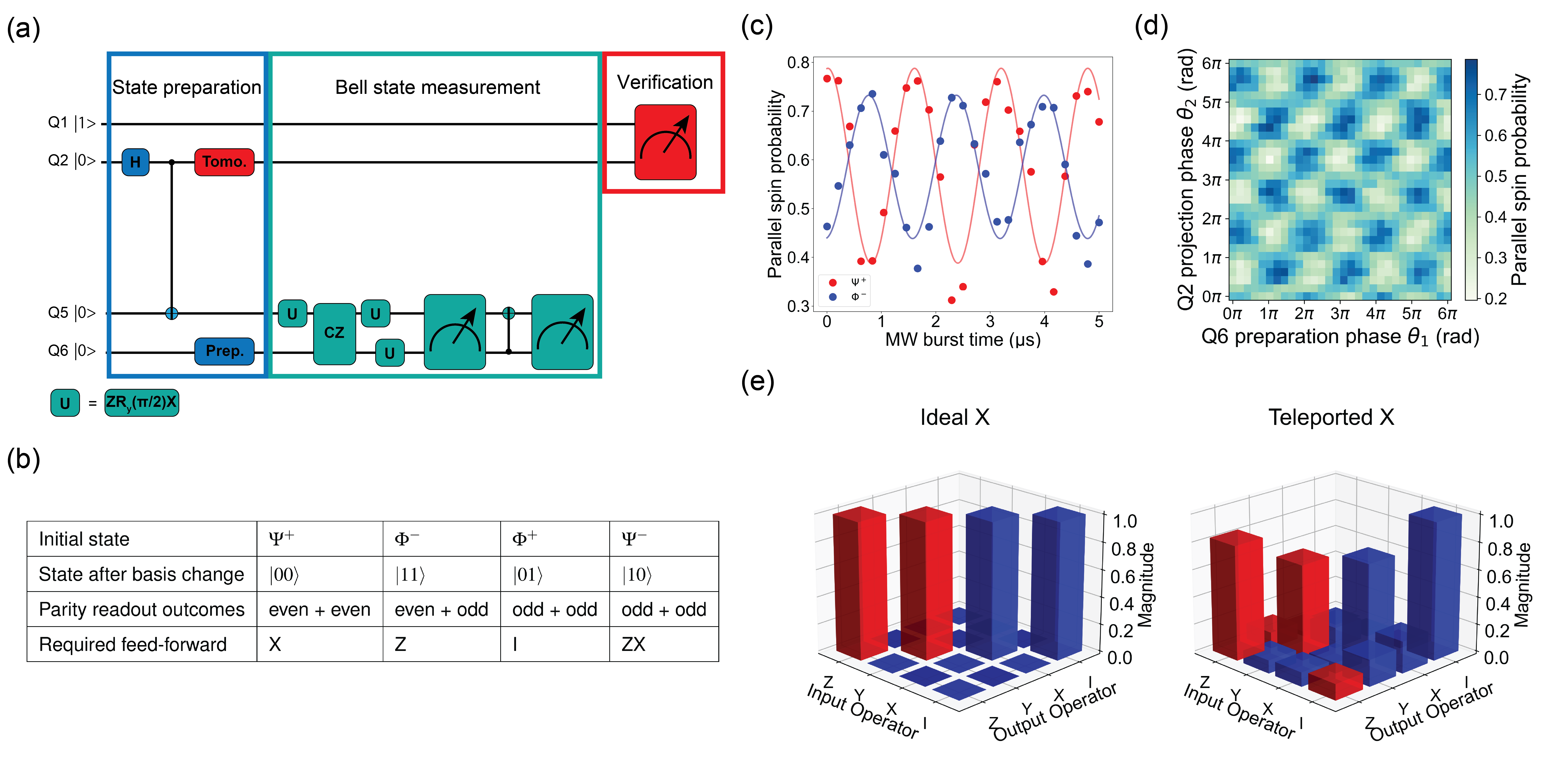}
\caption{\label{fig:fig4}\textbf{Quantum state teleportation.}
a) Quantum circuit for teleportation using mobile spin qubits, divided into state preparation (blue), Bell state measurement (green), and verification (red) sections. The U gate represents the composite gate Z$R_y (\pi/2)X$. See main text for more details.
b) Look-up table relating the input states to the computational basis states they are mapped to by the basis transformation, the parity readout outcomes given the practical constraints of successive parity measurements, and the required feed-forward operations to complete teleportation.
c) Parallel spin probability measured on Q1Q2 as a function of the microwave burst time on Q6, showing Rabi oscillations on Q2 after teleportation, post-selected on detecting $\Psi^{+}$ (blue) and $\Phi^{-}-$ (red) Bell states.
d) Two-dimensional map of the parallel spin probability measured on Q1Q2 after teleporting a superposition input state, post-selected on detecting the $\Psi^{+}$ Bell state.
e) Ideal and experimentally measured Pauli transfer matrices for the $X$ operation, obtained through quantum process tomography of the teleportation channel. Red (blue) bars represent positive (negative) values of the magnitude shown.
}
\end{figure*}
\subsection{\label{subsec:Tele}Quantum state teleportation}

Finally, to demonstrate that we can take advantage of spatially separated entangled spins created by the mobile spin-qubit approach, we implement a conditional post-selected quantum state teleportation protocol.
Figure 4(a) illustrates the quantum circuit for teleportation using mobile spin qubits. The circuit is divided into three main sections: preparation of Q6 and Bell-state preparation of Q2 and Q5 through the controlled interaction of mobile spins (blue); a Bell-state measurement of Q5 and Q6 (green); and verification (red). 

In more detail, for Bell-state preparation we first perform single-qubit operations with Q2 and Q5 in dots 2 and 5, then shuttle them together to implement a CZ operation, and finally separate them to their original positions to create a distributed entangled state between distant qubits.

The Bell-state measurement is achieved through a series of operations: first, a measurement basis transformation is applied to map the four Bell states onto the four computational basis states ($\lvert$00⟩, $\lvert$01⟩, $\lvert$10⟩, $\lvert$11⟩) of Q5 and Q6. Then, two sequential parity readouts, separated by a CNOT gate, are performed. The CNOT between Q5 and Q6 is implemented by first applying a $R_y (\pi/2)$ gate to Q6, followed by a CZ operation achieved by pulsing a barrier gate between static dots to control the exchange interaction, and finally another -$R_y (\pi/2)$ gate to Q6.
It is important to note that due to the mixing of the triplet $T_{0}$ and the singlet $S$ during PSB parity readout~\cite{amanda_PSB_2021,philips_universal_2022}, while this method can distinguish between the $\lvert$00⟩ and $\lvert$11⟩ states, it cannot differentiate between the $\lvert$10⟩ and $\lvert$01⟩ states. This limitation impacts the ability to fully discriminate between all four Bell states, hence we here achieve conditional quantum state teleportation. Alternative measurement schemes will be necessary to overcome this constraint.
Figure 4(b) provides a guide for the required feedforward operations based on the Bell measurement outcome and the corresponding parity readout results. Due to the inability to distinguish states when the first parity readout is odd, we focus the analysis on post-selected cases where we detect either the $\Psi^{+}$ or $\Phi^{-}-$ Bell states. These cases correspond to applying X and Z gates, respectively, to the teleported spin state.

For verification, we apply a tomography pulse on Q2 followed by parity readout of Q2 against the reference qubit Q1. The application of tomography pulses to Q2 prior to the Bell measurement is justified by the quantum teleportation protocol's structure. Once the initial Bell pair is created between Q2 and Q5, the subsequent tomography pulses on Q2 and Bell measurement between Q5 and Q6 are independent events -- their order does not affect the teleported state. This approach is particularly advantageous in the implementation in this work where measurement times (40 $\mu$s) exceed coherence times, as it minimizes the delay between state preparation and the state projection by tomography pulses.
Similar to photonic teleportation experiments ~\cite{bouwmeester_photonic_1997,pirandola_teleport_2015}, we employ post-selection which, while reducing the overall success probability, preserves the quantum nature of the protocol.

Figure 4(c) illustrates that the spin polarization prepared via Rabi oscillations in Q6 is successfully transferred and observable in the state of Q2, as expected. We note that this step in itself only confirms that a classical probability was transferred, and provides only partial evidence for successful quantum state teleportation. The blue and red curves show the oscillations after post-selecting the Bell states $\Psi^{+}$ and $\Phi^{-}$, respectively. In this case, no tomography pulses are applied. 

Next, we teleport superposition states prepared with different phases in Q6. This is achieved by applying an $R_x (\pi/2)$ pulse followed by an $R_Z$ ($\theta_1$) rotation on Q6 during state preparation. Post-selecting on the $\Psi^{+}$ Bell state, the phase information is transferred to Q2 where we apply an $R_{Z}$($\theta_2$) rotation followed by another $R_x (\pi/2)$ pulse, which projects the phase information onto the Z-basis measurement. Figure 4d shows the resulting oscillations, which depend on both the $R_Z$ ($\theta_1$) operation applied to Q6 and on the $R_Z$ ($\theta_2$) rotation on Q2. These oscillations confirm that the phase information is preserved and accurately transferred through the teleportation process. Nevertheless, similar to the data in panel c, this is only partial evidence of successful quantum state teleportation.

Following this, quantum process tomography (QPT)~\cite{ariano_qpt_2001} is performed to quantitatively analyze the fidelity of the teleportation process, such that we can assess whether genuine quantum state teleportation was achieved by verifying that the process fidelity exceeds the classical limit of 2/3. We focus on the post-selected cases corresponding to the $\Psi^+$ Bell state measurement outcome, which implements an effective $X$ gate operation on the teleported state.
Figure 4(e) present the ideal (left) and experimental (right) Pauli transfer matrices for this $X$ operation, corresponding to the even+even ($|00\rangle$) parity readout outcomes between Q5 and Q6. The experimental Pauli transfer matrices are reconstructed using least-squares optimization with complete positivity and trace-preserving (CPTP) constraints to ensure the physical validity of the estimated quantum process~\cite{mg_projectedleast_2020,Surawy_projectedleast_2022}. After correcting for SPAM errors, we determine that the average fidelity of the $X$ gate is $86.7 \pm 0.9\%$, which corresponds to the fidelity of the teleportation process. The uncertainty is the standard deviation obtained through bootstrap resampling. This value significantly exceeds the classical bound of 2/3, demonstrating that this protocol achieves genuine quantum state teleportation.

In analyzing the sources of imperfection in the system, the experimental infidelities can be primarily attributed to several factors in order of significance. The most substantial contribution comes from the errors in the Q5-Q6 Bell state measurement, which was created using a local CZ gate without optimized exchange coupling. This directly affects the transformation from the Bell basis to the ZZ basis, as well as the CNOT gate in between two parity readouts. Next in importance are the errors in the Q2-Q5 Bell state preparation, primarily limited by imperfect initialization of the individual qubits. Following these, additional errors are introduced through the parity readout process (see supplementary information).

\section{\label{sec:conclusion}Conclusion}
In this work, we demonstrate the concept of two-qubit gates between mobile semiconductor spin qubits. 
The exchange interaction is activated and precisely controlled in time by moving two electrons towards each other in the minima of two traveling conveyor potentials, which allows for a two-qubit CZ gate fidelity of about 99\%. Interestingly, we observe a saturation of the exchange coupling and enhanced dephasing times in elongated conveyor potential minima, possibly due to the formation of strongly correlated electron states. While further investigation is needed to confirm the microscopic nature of these states, this operating regime could provide an interesting avenue for implementing robust two-qubit gates.

Moreover, we showcase the potential of distributed quantum computing through conditional post-selected quantum state teleportation between distant qubits. In future quantum processors, crucial operations such as magic-state distillation could be performed in dedicated regions and the resulting states could be distributed to computational zones through teleportation, enabling efficient resource sharing across the processor. The next step will be to demonstrate deterministic teleportation by incorporating fast, non-demolition readout techniques that enable real-time feed-forward operations based on four distinguishable Bell state measurement outcomes~\cite{barrett_teleport2004,Chou_teleport2018,Takeda_teleport2013}.

Scaling up quantum processors based on mobile spin qubits would require advances along three key directions. First, increasing the range of qubit transport through long-distance conveyor channels with shared control lines~\cite{taylor_shuttle_2005,boter_spyder_2022,xue_sisige_2024}. Second, as illustrated in Figure 1(a), implementing arrays of storage zones connected by shared conveyor belts to exploiting flexible connectivity between many qubits. Third, realizing simultaneous shuttling operations in parallel conveyor channels to increase the number of mobile qubits that can be controlled independently. These advances, building upon the demonstration of high-fidelity two-qubit operations and quantum state teleportation with mobile spin qubits shown here, represent vital steps toward large-scale, reconfigurable quantum processors.

\section*{\label{sec:ackn}Acknowledgments}
We wish to thank S.G.J. Philips for writing control libraries and designing the PCB, R. Schouten, R. Vermeulen, O. Benningshof and T. Orton for support with the measurement setup and dilution refrigerator, B. Undseth for discussions on scalable architectures, X. Xue for discussions on tuning of two-qubit gates, C. Wang for discussions on benchmarking, and other members of the Vandersypen, Veldhorst, Scappucci, and Dobrovitski groups for fruitful discussions.
We acknowledge financial support from the Army Research Office (ARO) under grant number W911NF2310110. M. Rimbach-Russ acknowledges support from the Netherlands Organization of Scientific
Research (NWO) under Veni Grant No. VI.Veni.212.223 and by the EU through H2024 QLSI2. The views and conclusions contained in this document are those of the authors and should not be interpreted as representing the official policies, either expressed or implied, of the ARO or the US Government. The US Government is authorized to reproduce and distribute reprints for government purposes notwithstanding any copyright notation herein. 
% \end{acknowledgments}

\section*{\label{sec:contrib}Author contributions}
Y.M. and M.D.S. performed the experiments and data analysis. Simulations were carried out by Y.M. Libraries for experimental control were written by S.L.S. and Y.M. Y.M., M.D.S., M.R.R. and L.M.K.V. contributed to data interpretation. L.T. fabricated the device, while S.V.A. refined the device design. A.S. and G.S designed and grew the heterostructure. Y.M., M.D.S., and L.M.K.V. wrote the manuscript with comments by all authors. Y.M conceived the project. L.M.K.V. supervised the project.

\section*{\label{sec:interests}Competing interests}
The authors declare no competing interests.

\section*{\label{sec:contrib}Data availability}
The raw measurement data and the analysis supporting the findings of this work are available on a Zenodo repository (https://doi.org/10.5281/zenodo.15052065).

\section*{\label{sec:methods}Methods}
\subsection{\label{supp:subsec:parity} Parity readout and initialization sequence}

Here we describe the details of the initialization sequence. To optimize the readout visibility through Pauli spin blockade, we utilize the (3,1)–(4,0) charge transition for dots 1-2 and the (1,3)–(0,4) transition for dots 5-6. First, we check the parity of the (Q1,Q2) and (Q5,Q6) pairs. If the parity is even, we apply an X gate to Q2 or Q5 using microwave-driven feedback control in the (3,1) and (1,3) charge states, respectively. We then remeasure the parity of the (Q1,Q2) and (Q5,Q6) pairs and post-select for odd parity. Subsequently, we adiabatically transition the (Q1,Q2) and (Q5,Q6) pairs from (4,0) to (3,1) and from (0,4) to (1,3) charge states, respectively, over a duration of 50 ns. To enhance adiabaticity during this process, we apply pulses of approximately +200 mV to the inter-dot barrier gates (B1 and B5) compared to the readout conditions. When this adiabatic initialization is successful, the spin states of (Q1,Q2) and (Q5,Q6) become (1,0) and (1,0), respectively, as determined by the Zeeman energy differences between each pair. Finally, we apply an X gate to Q5, initializing the spin states of (Q1,Q2,Q5,Q6) to (1,0,0,0).
\subsection{\label{supp:subsec:twotone} Two tune conveyor pulse}
Following previous work~\cite{desmet_highfidelityshuttle_2024}, we use a combination of two sine waves for each gate: one at frequency $f$ and another at $f$/2, which can be expressed as $$V_n(t) = V^{DC}_n + \frac{A}{2}[\sin(2\pi ft - \phi_n) + \sin(\pi ft - \theta_n)],$$ where $V^{DC}_n$ represents the individual pulse offset, A is the amplitude, and $\phi_n$ and $\theta_n$ are the phase offsets for the respective frequency components. This approach creates wider potential barriers between neighboring conveyor minima, which significantly reduces the probability of charge leakage during transport.%, particularly in the presence of background disorder.
\subsection{\label{supp:subsec:RB_analysis} Interleaved randomized benchmarking}
We analyze the average gate fidelity of the shuttling based CZ gate using interleaved randomized benchmarking. In our analysis, the sequence fidelity \( F_t(L) \) was measured as \( P_{\downarrow\downarrow}(L) \), which is the probability that both qubits are in the spin-down state after applying \( L \) Clifford gates. The measured sequence fidelity \( F_t(L) \) is modeled as:

\begin{equation}
F_t(L) = A_t \cdot p_t^L + B_t
\end{equation}

where \( A_t \) and \( B_t \) are constants accounting for state preparation, measurement errors, and any residual offsets, while \( p_t \) is the depolarizing parameter. The fidelity of the Clifford gates \( F_C \) is then calculated from \( p_t \) using:

\begin{equation}
F_C = \frac{1 + 3p_t}{4}
\end{equation}

For the CZ gate fidelity \( F_{CZ} \), the sequence fidelity is measured with CZ gates interleaved between Clifford gates. The CZ gate fidelity is then estimated by comparing the decay rates of the sequence fidelities with and without the interleaved CZ gates:

\begin{equation}
F_{CZ} = \frac{1 + 3p_{CZ}/p_{\text{ref}}}{4}
\end{equation}

where \( p_{\text{ref}} \) is the depolarizing parameter obtained from the reference (non-interleaved) benchmarking sequence.
\bibliographystyle{naturemag}
% \bibliography{references}% Produces the bibliography
\bibliography{manualbib}

% \appendix
\section*{\label{sec:Supp}Supplementary information}
\begin{figure*}[ht]
\centering
\includegraphics[width=0.88\textwidth]{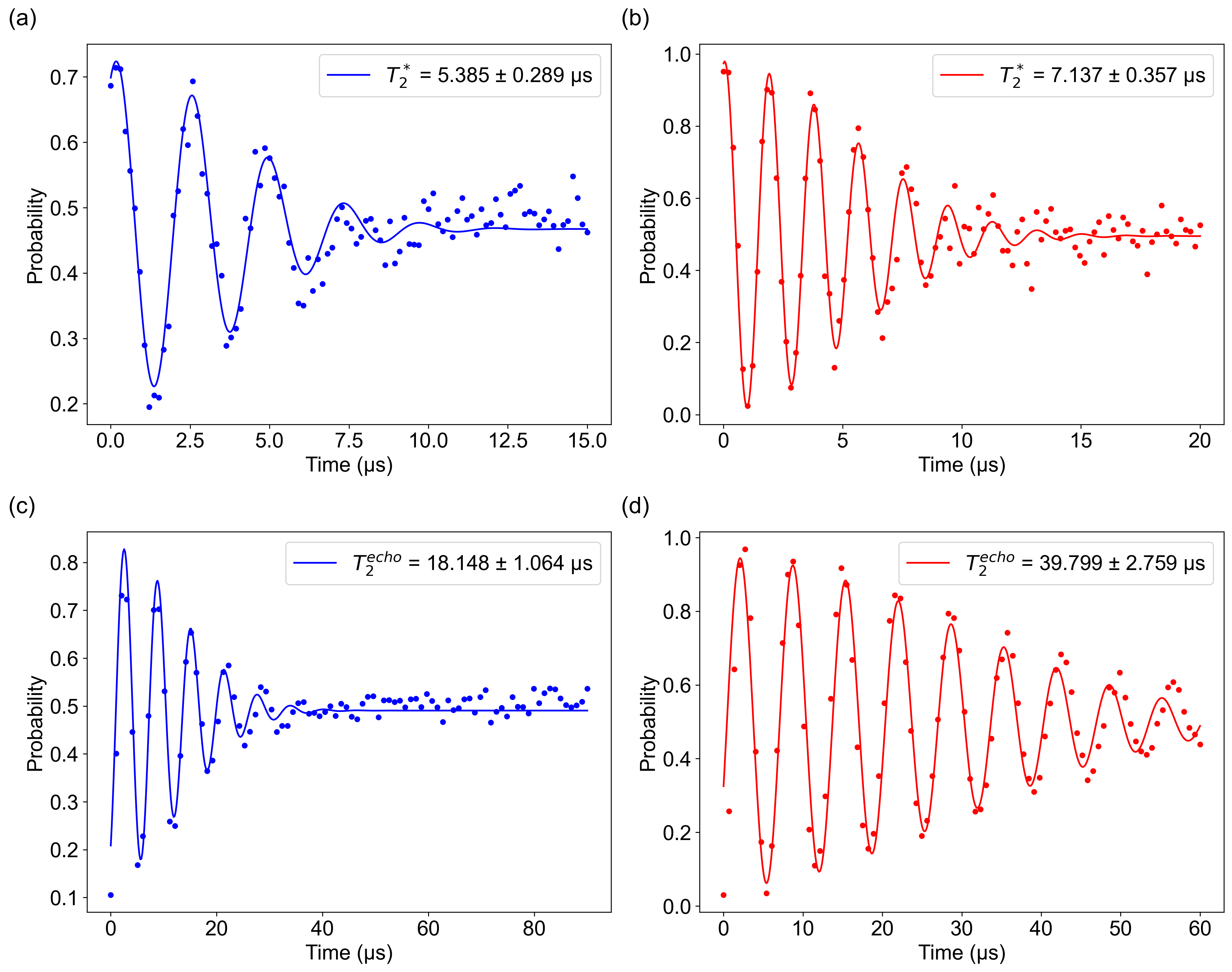}
\caption{\label{supp:fig:RE}\textbf{Ramsey and Echo measurement of Q2 and Q5} a,b) Ramsey interference measurements showing free induction decay for Q2 and Q5, yielding dephasing times $T_{2}^{*}$ of 5.385 $\pm$ 0.289 $\mu$s and 7.137 $\pm$ 0.357 $\mu$s, respectively. The solid lines represent fits to the data using Gaussian-decayed sinusoidal functions of the form $A\exp(-(t/T_2^*)^2)\sin(2\pi ft + \phi) + C$. c,d) Hahn echo measurements for Q2 and Q5, demonstrating extended coherence times $T_{2}^{\text{echo}}$ of 18.146 $\pm$ 1.064 $\mu$s and 39.799 $\pm$ 2.759 $\mu$s, respectively. The solid lines show the Gaussian-decayed sinusoidal fits to the echo data. The error bars represent one standard deviation extracted from the fit.
}
\end{figure*}

%\subsection{\label{supp:subsec:valley} Valley splitting}

%Magnetospectroscopy was employed to measure the two-electron singlet-triplet energy splitting, $E_{ST}$, in this device as reported in ~\cite{degli_esposti_low_2024}. Although these measurements were carried out under a different cooldown and gate voltage configuration compared to our current shuttling experiments, the estimated $E_{ST}$ values for all quantum dots are provided in Table \ref{Supp:tab:valley}. In strongly confined quantum dots, $E_{ST}$ serves as a conservative lower bound for the single-particle valley splitting $E_{v}$ ~\cite{ercan_strong_2021}. For our shuttling experiments, the singlet-triplet energy splitting in each quantum dot far exceeds the Zeeman splitting, thereby suppressing dephasing at spin valley hot-spots.

\begin{figure*}[h]
\centering
\includegraphics[width=0.76\textwidth]{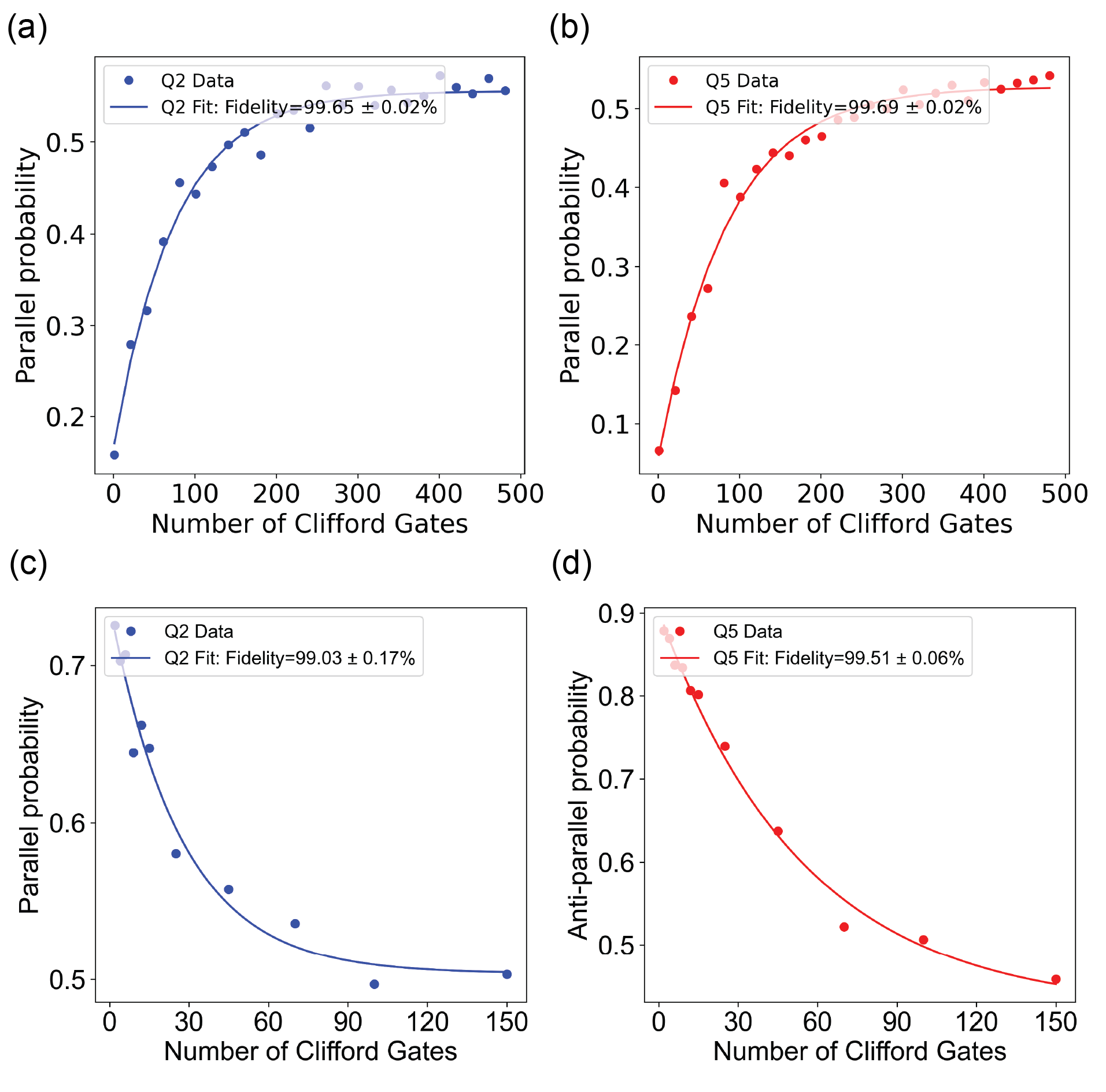}
\caption{\label{supp:fig:RB}\textbf{Randomized benchmarking of Q2 and Q5}  a) Randomized benchmarking data for Q2  showing the parallel spin probability as a function of the number of Clifford gates. The solid line shows a fit yielding an average single-qubit gate fidelity of 99.65 $\pm$ 0.02\%. b) Same measurement for Q5 yielding an average single-qubit gate fidelity of 99.69 $\pm$ 0.02\%. c) Randomized benchmarking with simultaneous driving of Q2 and Q5, showing reduced fidelity for Q2 at 99.03 $\pm$ 0.17\%. The degradation is likely due to heating-induced frequency shifts of the qubit~\cite{undseth_hotter_2023}. d) Simultaneous driving measurement for Q5 showing fidelity of 99.51 $\pm$ 0.06\%, which exhibits less degradation compared to Q2. The overall joint fidelity for simultaneous operation is 98.54 $\pm$ 0.18\%. The error bars represent one standard deviation extracted from the fit.}
\end{figure*}

%\begin{table}[h]
%\centering
%\caption{$E_{ST}$, as reported by ~\cite{degli_esposti_low_2024}, for each quantum dot in this device, which serves as a lower bound for the single-particle valley splitting.}
%\label{Supp:tab:valley}
%\begin{tabular}{||c c c c c c||}
% \hline
% QD 1 & QD 2 & QD 3 & QD 4 & QD 5 & QD 6 \\ [0.5ex]
% \hline\hline
% \SI{208}{\micro eV} & \SI{174}{\micro eV} & \SI{276}{\micro eV} & \SI{208}{\micro eV} & \SI{243}{\micro eV} & \SI{278}{\micro eV}\\  [1ex] 
% \hline
%\end{tabular}
%\end{table}

\subsection{\label{supp:subsec:device} Device fabrication}
    
In this work, the device is fabricated on a \ch{^{28}Si}/SiGe heterostructure ~\cite{lawrie_quantum_2020}. Initially, a \SI{1.5}{\micro \meter} thick, linearly graded \ch{Si_{1-x}Ge_x} buffer layer is deposited on a silicon wafer, which is then capped with a \SI{300}{\nano \meter}-thick relaxed \ch{Si_{0.7}Ge_{0.3}} spacer. Next, a \SI{7}{\nano \meter}-thick, tensile-strained \ch{^{28}Si} quantum well enriched to 800 ppm is grown ~\cite{degli_esposti_low_2024}. A \SI{30}{\nano \meter} thick \ch{Si_{0.7}Ge_{0.3}} spacer, passivated with dichlorosilane at 500$^\circ$C ~\cite{degli_esposti_wafer-scale_2022}, is employed to separate the quantum well from the gate stack. Ohmic contacts to the two-dimensional electron gas within the quantum well are then formed via phosphorus ion implantation. Following this, a \SI{10}{\nano \meter} \ch{Al2O3} layer is deposited, succeeded by three Ti:Pd layers (with thicknesses of 3:17, 3:27, and 3:37 \SI{}{\nano \meter}) deposited by electron beam evaporation; each metallic layer is interleaved with a \SI{5}{\nano \meter} \ch{Al2O3} film grown by atomic layer deposition. Finally, an extra \SI{5}{\nano \meter} \ch{Al2O3} layer is added atop the gate stack, upon which a 5:200 \SI{}{\nano \meter} Ti:Co micromagnet is deposited to enable qubit control via magnetic field gradients.

\begin{figure*}[ht]
\centering
\includegraphics[width=0.75\textwidth]{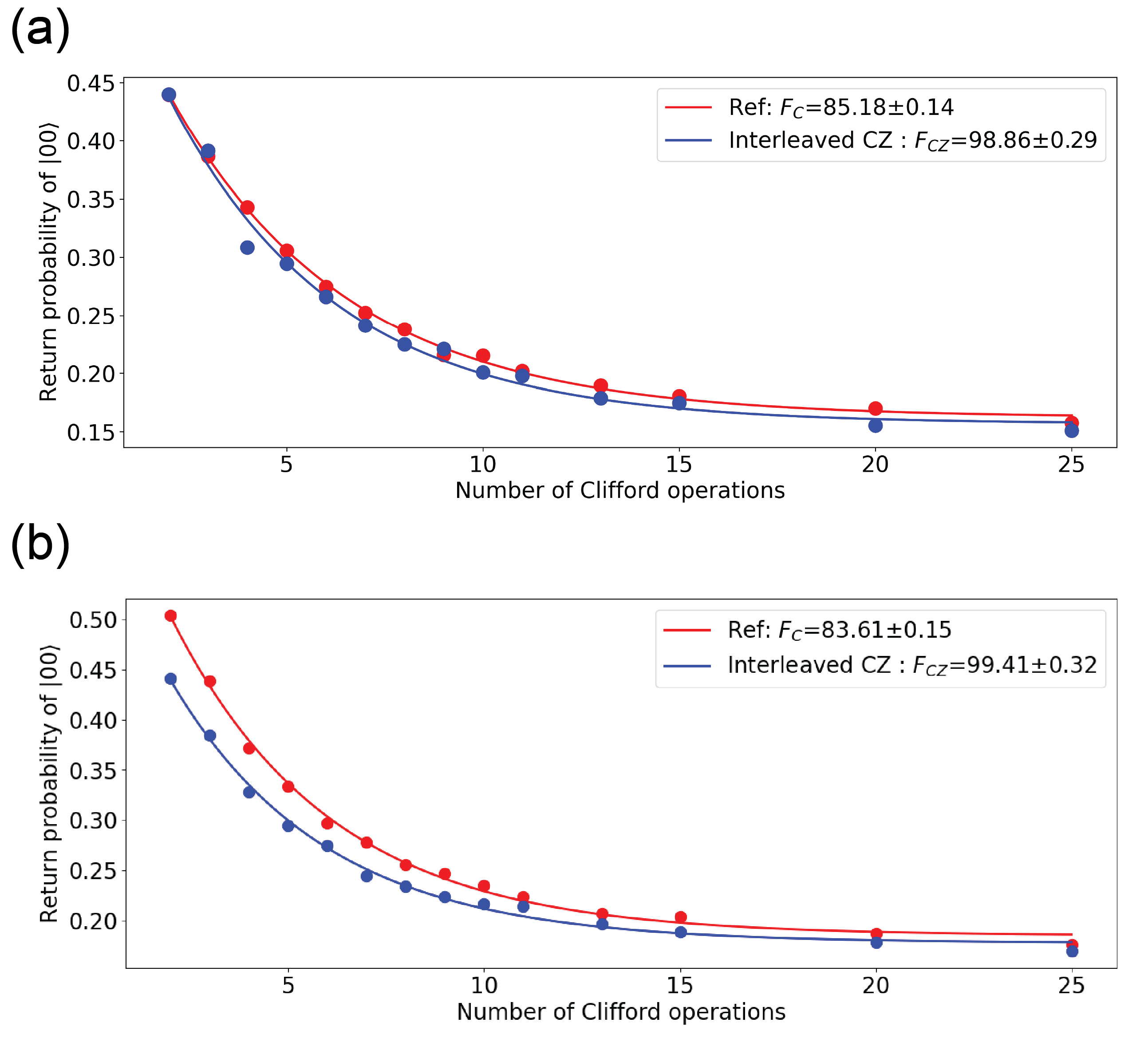}
\caption{\label{supp:fig:IRB}\textbf{Statistics of interleaved randomized benchmarking measurements.} a) First set of measurements showing sequence fidelity versus number of Clifford operations for both reference (red) and interleaved CZ gate (blue) sequences. The measured average Clifford gate fidelity is $F_C = 85.18 \pm 0.14\%$ and the extracted CZ gate fidelity is $F_{CZ} = 98.86 \pm 0.29\%$. b) Second set of measurements showing sequence fidelity versus number of Clifford operations for both reference (red) and interleaved CZ gate (blue) sequences, yielding an average Clifford gate fidelity of $F_C = 83.61 \pm 0.15\%$ and a CZ gate fidelity of $F_{CZ} = 99.41 \pm 0.32\%$. This dataset is not included in the main manuscript due to the significant amplitude difference between the two decay curves, which likely results from low-frequency drift affecting single-qubit gate errors and state preparation and measurement (SPAM) errors over the course of the measurement. The error bars represent one standard deviation extracted from bootstrap resampling.}
\end{figure*}

\begin{table}[ht]
\caption{Conveyor mode control parameters for Fig. 2(b).}
\begin{tabular}{lccc}
\hline
Gate & Amplitude (mV) & DC Offset (mV) & Phase Offset $\phi$/$\theta$ ($2\pi$) \\
\hline
VP1 & 0 & 0 & -0.7/0.0 \\
VB1 & 0 & -80 & 0.0/0.05 \\
VP2 & 120 & -90 & 0.1/0.3 \\
B2 & 120 & 140 & 0.6/0.55 \\
P3 & 120 & -120 & 1.1/0.8 \\
B3 & 100 & $V_{B3}$ & 1.6/1.05 \\
P4 & 120 & 85 & 1.1/0.8 \\
B4 & 120 & 140 & 0.6/0.55 \\
P5 & 120 & -120 & 0.1/0.3 \\
B5 & 0 & -90 & 0.0/0.0 \\
P6 & 0 & 50 & 0.0/0.0 \\
B6 & 0 & -30 & 0.0/0.0 \\
\hline
\end{tabular}
\label{tab:conveyor_params}
\end{table}
\begin{table}[ht]
\caption{Conveyor mode control parameters for high fidelity CZ operations (Fig. 2(d) and Fig. 3. Here the voltage on P3 and P4 are finely tuned to improve the charge symmetry between Q2 and Q5}.
\begin{tabular}{lccc}
\hline
Gate & Amplitude (mV) & DC Offset (mV) & Phase Offset $\phi$/$\theta$ ($2\pi$) \\
\hline
VP1 & 0 & 0 & -0.7/0.0 \\
VB1 & 0 & -80 & 0.0/0.05 \\
VP2 & 120 & -90 & 0.1/0.3 \\
B2 & 120 & 140 & 0.6/0.55 \\
P3 & 120 & -110 & 1.1/0.8 \\
B3 & 100 & 9.5 & 1.6/1.05 \\
P4 & 120 & 95 & 1.1/0.8 \\
B4 & 120 & 140 & 0.6/0.55 \\
P5 & 120 & -120 & 0.1/0.3 \\
B5 & 0 & -90 & 0.0/0.0 \\
P6 & 0 & 50 & 0.0/0.0 \\
B6 & 0 & -30 & 0.0/0.0 \\
\hline
\end{tabular}
\label{tab:conveyor_params}
\end{table}
\begin{table}[ht]
\caption{Conveyor mode control parameters for merging two elongated moving potentials.}
\centering
\begin{tabular}{lccc}
\hline
Gate & Amplitude (mV) & DC Offset (mV) & Phase Offset $\theta$ ($2\pi$) \\
\hline
VP1 & 0 & 0 & 0.0 \\
VB1 & 0 & -80 & 0.05 \\
VP2 & 160 & -90 & 0.3 \\
B2 & 160 & 140 & 0.55 \\
P3 & 160 & -110 & 0.8 \\
B3 & 160 & -10 & 1.05 \\
P4 & 160 & 80 & 0.8 \\
B4 & 160 & 140 & 0.55 \\
P5 & 160 & -120 & 0.3 \\
B5 & 0 & -90 & 0.0 \\
P6 & 0 & 50 & 0.0 \\
B6 & 0 & -30 & 0.0 \\
\hline
\end{tabular}
\label{tab:conveyor_merge_conditions}
\end{table}

\begin{figure*}[ht]
\centering
\includegraphics[width=1\textwidth]{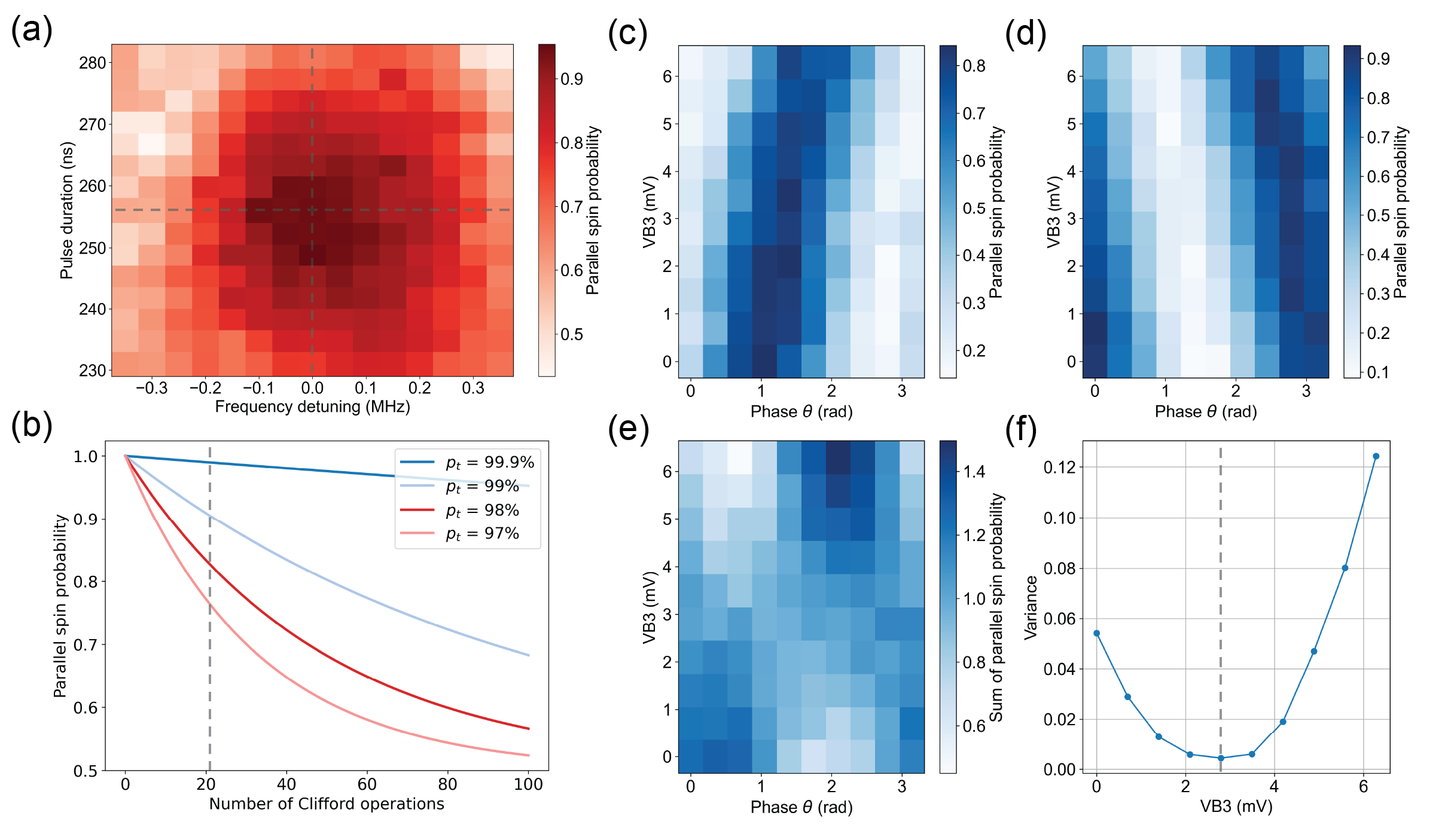}
\caption{\label{supp:fig:RB}\textbf{Single- and two-qubit gate optimization.} a) Two-dimensional scan of single-qubit gate calibration for Q5, showing the sequence fidelity of the RB with 20 Clifford gates as a function of microwave frequency detuning and $R_x (\pi/2)$ rotation pulse duration. The initial state is prepared in the excited state to avoid false optimization at fully off-resonant conditions. The dashed lines indicate the calibrated value of parameters. Similar optimization procedures were also performed for the CZ gate after the initial tuning shown in panels (c-f). b) shows the simulated parallel spin probability as a function of the number of Clifford gates in a RB sequence under different depolarizing factors $p_t$. It demonstrates why maximizing the parallel spin probability for a fixed number of Clifford gates leads to improved fidelity. c,d) Controlled-phase rotation angle of Q5 as a function of barrier gate B3 AC voltage offset, with control qubit Q2 prepared in $|0\rangle$ and $|1\rangle$ states respectively, for a total gate time of 54 ns. e) Combined data from (c) and (d), where the optimal CZ gate condition corresponds to a phase evolution that is independent of the B3 voltage (horizontal stripe pattern). f) Variance of the phase evolution along the x-axis in (e). The minimum indicates the optimal B3 voltage offset for implementing the CZ gate at the given gate duration~\cite{mills_high-fidelity_2022}. The solid line only connects the datapoint and is a guide to the eye.
The error bars represent one standard deviation extracted from bootstrap resampling.}
\end{figure*}

\begin{figure*}[ht]
\centering
\includegraphics[width=0.98\textwidth]{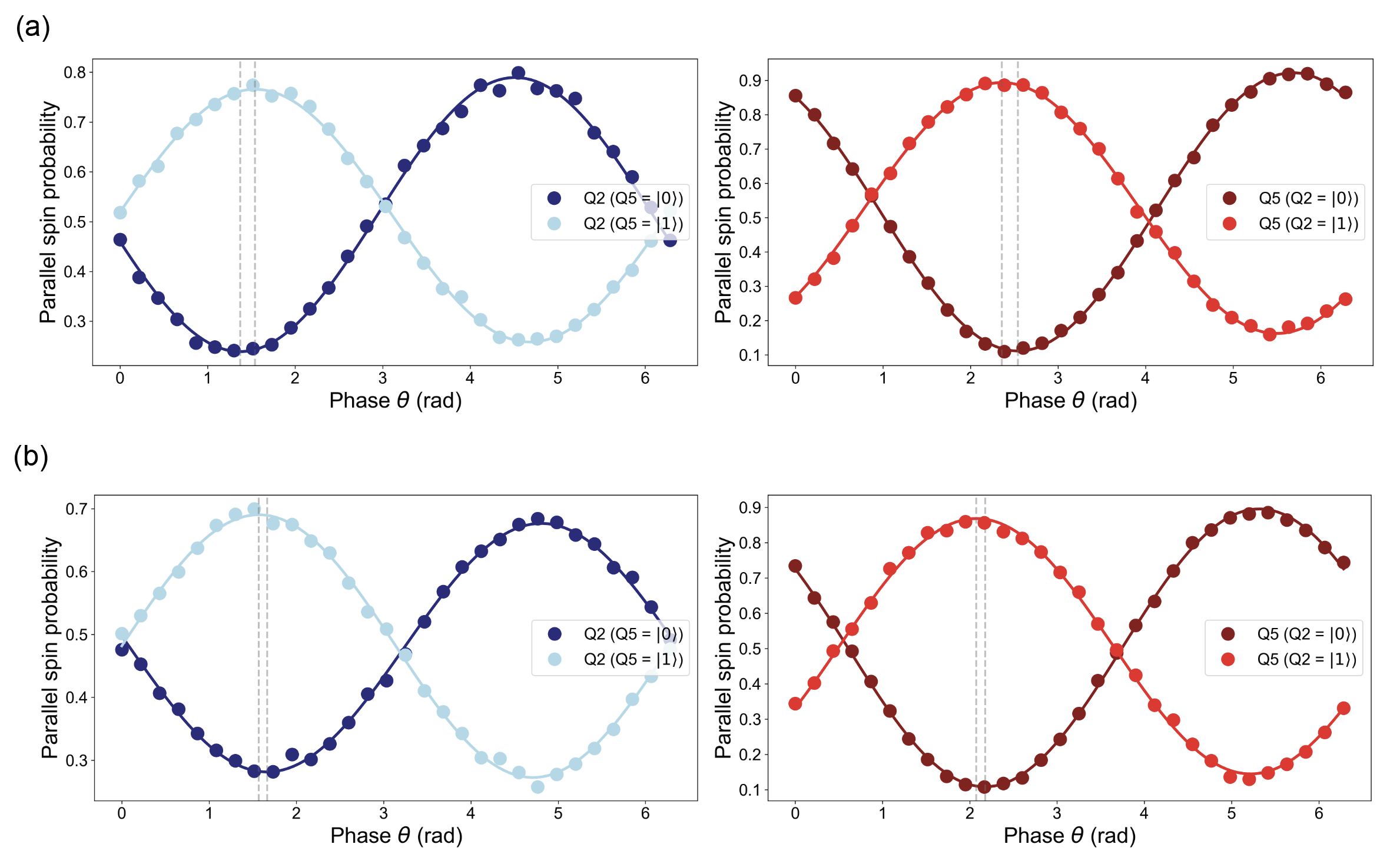}
\caption{\label{supp:fig:heating}\textbf{Heating effect on CZ calibration} Heating effect on CZ calibration. a) CZ calibration scan without MW pre-burst. b) CZ calibration scan with an  off-resonant 8 $\mu$s MW pre-burst applied before qubit initialization. For both panels, after initialization, we measure the parallel spin probability using a sequence where one spin (Q2 for left panel, Q5 for right panel) is prepared in a superposition state using an Rx($\pi$/2) gate while the other spin is initialized in either $|0\rangle$ or $|1\rangle$. The presence of the MW pre-burst significantly alters the phase evolution during the shuttling based CZ operation, illustrating how MW-induced heating can affect calibration accuracy. This effect may be caused by MW-induced shifts in qubit frequencies, which impact the phase acquisition during the operation~\cite{undseth_hotter_2023}. This effect, rather than imperfect CZ operations, explains the slight deviation from a $\pi$ controlled phase observed in the main text Figure 3c.
}
\end{figure*}
\clearpage

\subsection{\label{supp:subsec:CZ} Quantitative analysis of error source in CZ gate operation}

We analyze the dephasing effects during our CZ gate operation by considering low-frequency fluctuations  in the exchange interaction $J$ and individual qubit frequencies, causing random deviations between the actual operation $U_{exp}$ and ideal operation $U_{ideal}$. We model this as a stochastic unitary $U_{exp}$ dependent on noise parameter $x$. 
From the $T_2^*$ measurements at different conveyor periods shown in Figure 2d, we observe $T_2^*$ values ranging from approximately 0.2-4.0 $\mu$s during the interaction. We make the following assumption for these fluctuations. The fluctuations are Gaussian distributed with zero mean, stationary , and described by a $1/f$ power spectral density. Furthermore, we assume that the noise only gives rise to an accumulated phase, thus, we can approximate this noise using quasistatic fluctuations. By integrating over the corresponding frequencies, we obtain $\sigma^2=2 \int_{t_{\mathrm{m}}^{-1}}^{t_{\mathrm{e}}^{-1}} \frac{S_x}{f} df$, where the effective standard deviation of the noise $\sigma$ is proportional to $1/T_2^*$. For a measurement time $t_m$ and gate operation time $t_e$, this relationship can be expressed as~\cite{1offnoise_2008,wang_operating_2024}:
$$\sigma \propto \frac{1}{T_2^*} \propto \sqrt{\ln \frac{0.401}{t_e / t_m}}$$
Here the total time $t_e$ of CZ gate is 58 ns. 
To estimate the gate infidelity, we compare two measurement time scales: the $T_2^*$ measurement time of 138 s and the IRB measurement time of 5160 s. Using the relationship above, we calculate the effective noise standard deviation $\sigma$ for the IRB scenario. We calculate the average gate fidelity between the experimental unitary and the ideal CZ gate unitary as:

$$F = \frac{\left\langle \left| \operatorname{tr}\left(U_{\text{ideal}}^{-1} U_{\exp}\right)\right|^{2} \right\rangle + d}{d(d+1)}$$

where the expectation value is given by~\cite{gatefidelityexpect_2019,wang_operating_2024}:

$$\left\langle \left| \operatorname{tr}\left(U_{\text{ideal}}^{-1} U_{\exp}\right)\right|^2 \right\rangle=\int_{-\infty}^{\infty}\left|\operatorname{tr}\left(U_{\text{ideal}}^{-1} U_{\exp}(x)\right)\right|^2 \frac{1}{\sqrt{2 \pi} \sigma} e^{-\frac{x^2}{2 \sigma^2}} dx$$

For the two-qubit system, $d = 4$, the experimental unitary matrix in the computational basis $|{\downarrow \downarrow}\rangle, |{\uparrow \downarrow}\rangle, |{\downarrow \uparrow}\rangle, |{\uparrow \uparrow}\rangle$ is given by:

$$ U_{\exp}(x) = \begin{pmatrix} 1 & 0 & 0 & 0 \\
0 & e^{-2\pi i\int f_1(t,x)dt} & 0 & 0 \\
0 & 0 & e^{-2\pi i\int f_2(t,x)dt} & 0 \\
0 & 0 & 0 & e^{-2\pi i\int (f_1(t,x) + f_3(t,x))dt} \end{pmatrix} $$

where $f_1(t,x) = f_{\mathrm{Q5}, \mathrm{Q2} \downarrow}(t,x)$, $f_2(t,x) = f_{\mathrm{Q2}, \mathrm{Q5} \downarrow}(t,x)$, $f_3(t,x) = f_{\mathrm{Q5}, \mathrm{Q2} \uparrow}(t,x)$.

By integrating the qubit frequencies $f_{Qi,Qj}(t,x)$ over time under the conveyor pulse used for CZ operation and the noise amplitude $x$, we estimate the total dephasing-induced infidelity during the CZ gate operation to be $\sim$0.22\%.

Additionaly, we evaluate the undesired residual SWAP processes during our CZ gate operation through numerical simulations based on a time-dependent Heisenberg Hamiltonian:

$$ H(t) = \frac{\Delta E_z}{2}(\sigma_1^z - \sigma_2^z) + J(t)(\sigma_1^x\sigma_2^x + \sigma_1^y\sigma_2^y + \sigma_1^z\sigma_2^z)/4 $$

where $\Delta E_z = 83$ MHz is the Zeeman energy difference between the qubits, and $J(t)$ follows the experimental pulse shape with a maximum value of 33 MHz. The first term represents the energy splitting between the $|01\rangle$ and $|10\rangle$ states, while the second term describes the exchange interaction between the spins.

We numerically solve the time-dependent Schrödinger equation using this Hamiltonian to obtain the unitary evolution operator over the complete gate sequence. The fidelity is evaluated by computing the average gate fidelity between the obtained unitary $U_{\exp}$ and the ideal operation $U_{\text{ideal}}$. For our experimental parameters, we find a contribution to the infidelity $1-F = 0.01\%$.

Although the ratio of $\Delta E_z/J \approx 2.5$ at the pulse peak is not sufficiently large to suppress SWAP errors, the actual gate infidelity remains very low. This follows directly from the expected pulse shape of $J$ that follows a smooth curve. The error from the SWAP process is $\sim S(-|\Delta E_z|)^2$, where $S(-|\Delta E_z|)$ is the spectral density of the exchange noise evaluated at the Zeeman energy difference. Remarkably, a linear ramping pulse translates to an almost Gaussian window that strongly suppresses SWAP processes~\cite{Rimbach-Russ_2023}.

The total estimated infidelity (0.22\% from dephasing and 0.01\% from imperfect adiabaticity) remains smaller than the experimentally observed infidelity of approximately 1.14\%. The remaining infidelity could possibly be attributed to extremely low-frequency discrete shifts in qubit parameters that might occur on timescales longer than the $T_2^*$ measurement time or thermal effects (heating) during operation that shift qubit frequencies~\cite{undseth_hotter_2023} and induce additional high-frequency noise in the device.

\begin{figure*}[ht]
\centering
\includegraphics[width=0.87\textwidth]{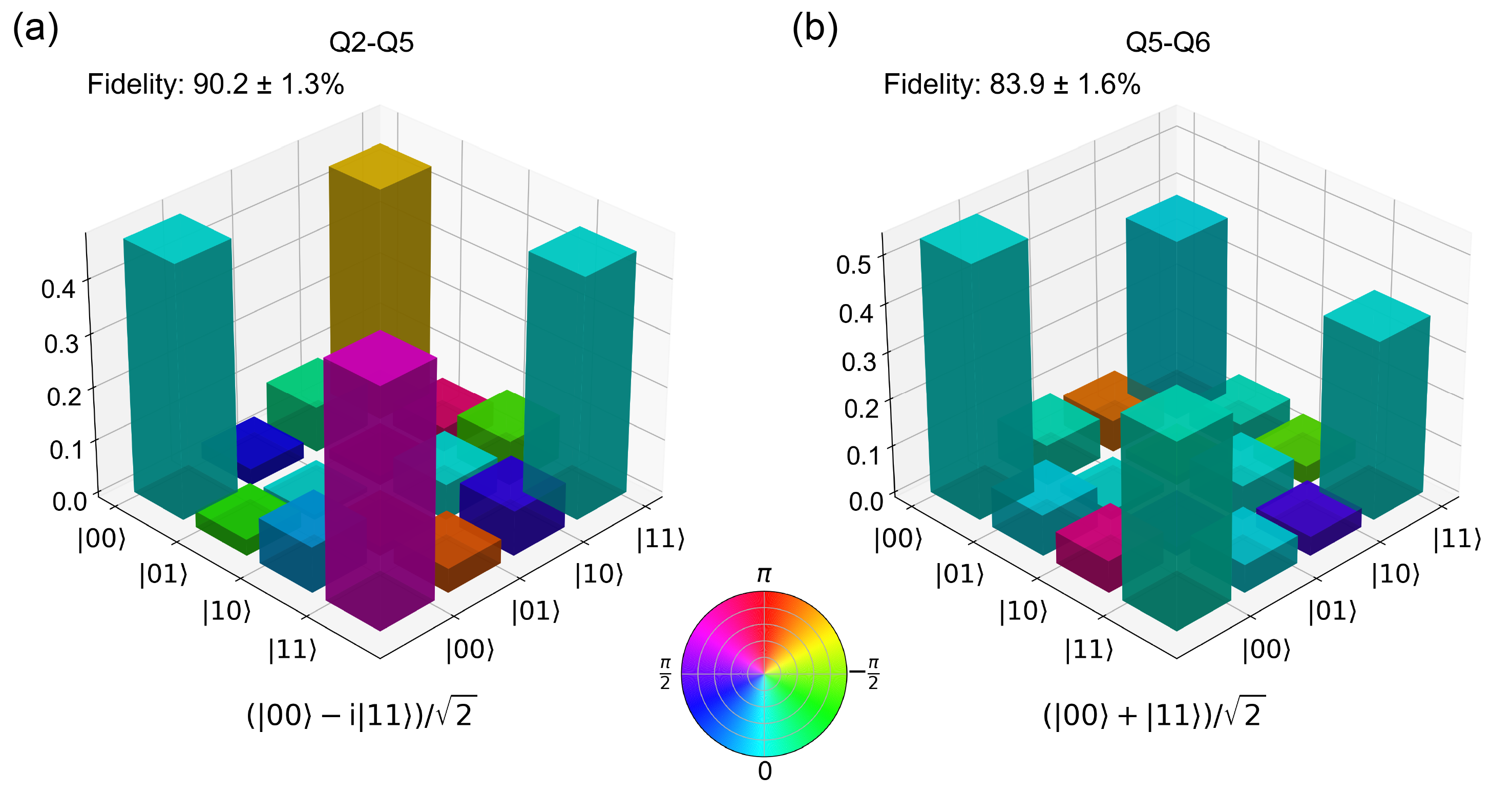}
\caption{\label{supp:fig:bell}\textbf{State tomography of entangled spin pairs used for teleportation.}  a) Reconstructed density matrix of the entangled state between mobile qubit Q2 and qubit Q5, prepared using shuttling-based controlled-phase operations. The basis states are ordered as {$\ket{00}$, $\ket{01}$, $\ket{10}$, $\ket{11}$}. We extract a Bell state fidelity of $90.2 \pm 1.3\%$ for this Q2–Q5 pair by fitting the phase of the reconstructed density matrix and calculating its overlap with the closest theoretical Bell state. b) Reconstructed density matrix of the entangled state between qubits Q5 and Q6, used as a resource for teleportation, with an extracted Bell state fidelity of $83.9 \pm 1.6\%$. Both density matrices were obtained through quantum state tomography with maximum likelihood estimation. Both real and imaginary components are represented by the bar heights, and the phase indicated by the color scale in radians.
}
\end{figure*}

\subsection{\label{supp:subsec:teleport} Quantitative analysis of error sources in quantum state teleportation}

We analyze the error sources in our teleportation protocol, beginning with characterizing the entangled Bell states through quantum state tomography. Fig.~\ref{supp:fig:bell}(a) shows the reconstructed density matrix for the entangled state between mobile qubit Q2 and qubit Q5, generated via shuttling-based CZ operations, while Fig.~\ref{supp:fig:bell}(b) presents the density matrix of the entangled state between qubit Q5 and qubit Q6, prepared using a local CZ gate in the static dots. From these measurements, we extract Bell state fidelities of $90.2 \pm 1.3\%$ for the Q2–Q5 pair and $83.9 \pm 1.6\%$ for the Q5–Q6 pair by fitting the phase of the reconstructed density matrices and calculating their overlap with the closest theoretical Bell states, obtained through quantum state tomography with maximum likelihood estimation. The uncertainties are the standard deviation obtained through bootstrap resampling.
Using the relation
$$
F_{\text{tele}} = \frac{1+2F_{\text{Bell}}}{3},
$$
we estimate that the error contributed by Bell state preparation for Q2–Q5 is $\sim$$6.5\pm 1.3\%$.

The Q5–Q6 pair, while characterized with a Bell state fidelity of 83.9\%, serves a different purpose in the protocol. These qubits are used for the Bell measurement basis transformation rather than directly contributing to the teleportation channel. The fidelity of operations on this pair affects the measurement process rather than directly transferring to teleportation infidelity in a straightforward manner.
The error analysis needs to consider the distinct roles of different qubit pairs in the teleportation protocol. While the Q2-Q5 Bell state directly impacts the teleportation channel quality with an estimated error contribution of $6.5 \pm 1.3\%$, the measurement process involving Q5-Q6 has a more complex relationship with the overall protocol fidelity.
This complexity was observed experimentally in the teleported Rabi oscillations (Figure 4(c)), where the amplitude varied significantly depending on the post-selected Bell state. When the $|\Phi^-\rangle$ state was post-selected, accurately estimating the process tomography matrix (PTM) through quantum process tomography (QPT) proved challenging, indicating that the error mechanisms manifest differently across the four Bell measurement outcomes.

We evaluate the average fidelity of the parity readout between Q5 and Q6 directly from the Rabi oscillation data. The return probability for the $|1\rangle$ state is measured to be $P_1 \approx 0.9799$, while the minimum probability (corresponding to the $|0\rangle$ state) is $P_0 \approx 1 - 0.0088 \approx 0.9912$. The average readout fidelity is then \(F = \frac{P_1 + P_0}{2} \approx 0.9856\), resulting in an average error per parity readout of \(\varepsilon \approx 0.0144\). Since two parity readouts contribute to the overall fidelity, the combined error from the parity readouts is roughly 2.9\%, though this estimation reflects the overall SPAM error, and the actual readout error is expected to be lower.

We note that the parity PSB readout error of Q1-Q2 used for teleportation verification is not part of the teleportation process itself, so we removed it using the following matrix: 
$$ \begin{pmatrix} 0.951 & 0.125 \\ 0.049 & 0.875 \end{pmatrix} \;. $$
The elements represent the probability of measuring each state given the prepared state: correctly measuring $|0\rangle$ when $|0\rangle$ was prepared (0.951), incorrectly measuring $|0\rangle$ when $|1\rangle$ was prepared (0.125), incorrectly measuring $|1\rangle$ when $|0\rangle$ was prepared (0.049), and correctly measuring $|1\rangle$ when $|1\rangle$ was prepared (0.875).
To remove the readout error from our data, we processed the probabilities as follows. If $P_{|1\rangle}$ is the probability of measuring $|1\rangle$ in our data, we calculated:
$$ \begin{pmatrix} P_{|0\rangle}^{\text{corrected}} \\ P_{|1\rangle}^{\text{corrected}} \end{pmatrix} = \begin{pmatrix} 0.951 & 0.125 \\ 0.049 & 0.875 \end{pmatrix}^{-1} \begin{pmatrix} 1 - P_{|1\rangle} \\ P_{|1\rangle} \end{pmatrix} \;.$$
Overall, while multiple error sources contribute to the protocol performance, the key areas for improvement include optimizing the Bell state preparation fidelity and enhancing the Bell measurement process in the static dots. Particularly, improving the local CZ gate performance between Q5 and Q6 would benefit the measurement basis transformation, which in turn would enhance the overall teleportation fidelity.

\begin{figure}[ht]
\centering
\includegraphics[width=0.4\textwidth]{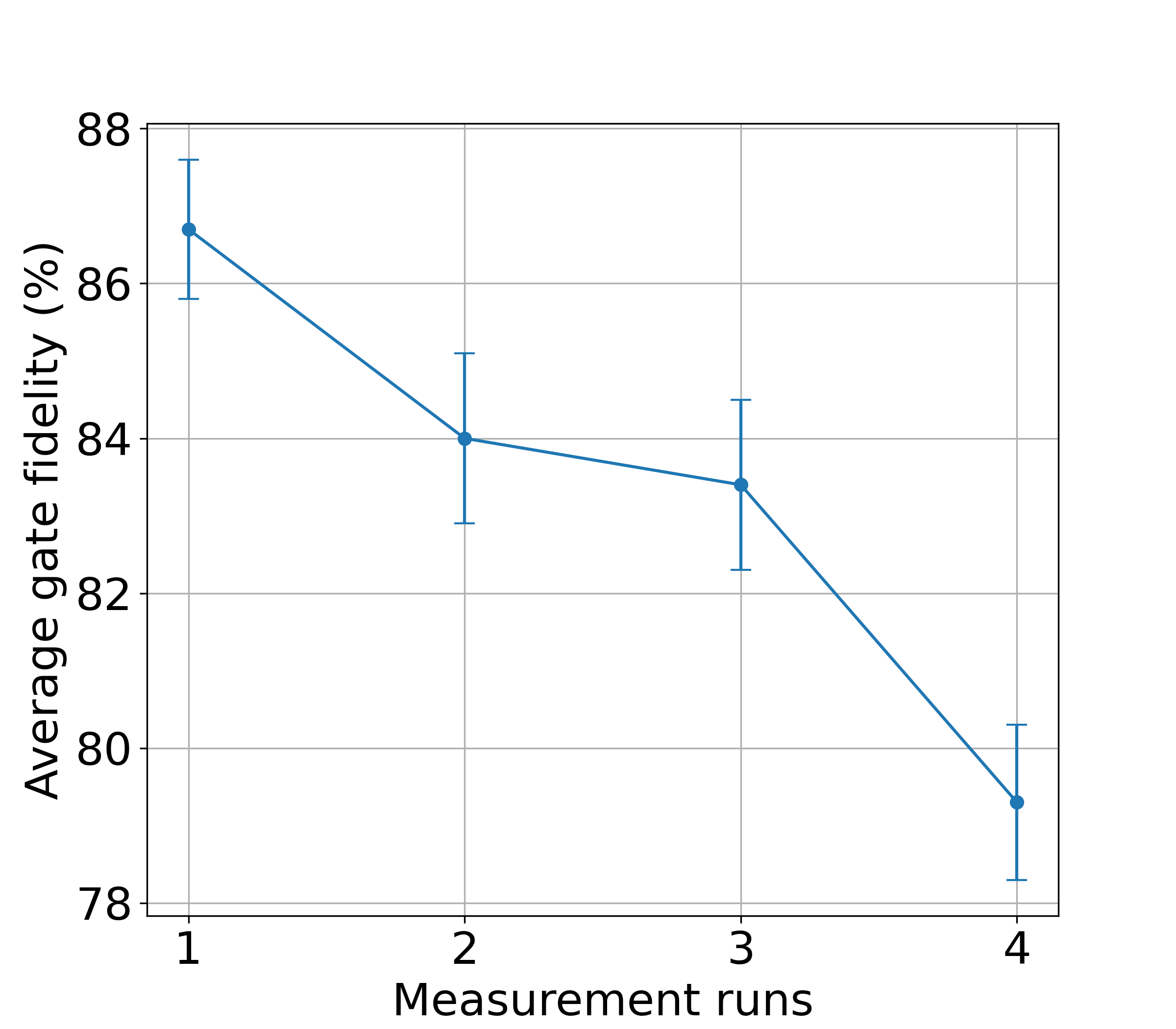}
\caption{\label{supp:fig:fidelities}\textbf{Average gate fidelities obtained from quantum process tomography (QPT) of the quantum teleportation protocol across four measurement runs.} The horizontal axis indicates the measurement run, while the vertical axis shows the average gate fidelity (\%). Error bars represent one standard deviation obtained from bootstrap resampling in each run. As time progresses, the initially tuned parameters drift, leading to a decrease in fidelity. Each runs takes approximately 7 minutes to complete.}

\end{figure}

Fig. ~\ref{supp:fig:fidelities} shows the average gate fidelities for four successive runs of the teleportation protocol without recalibrations in between. Each runs takes approximately 7 minutes to complete. As the measurements are repeated, parameter drift causes a gradual decrease in fidelity. Since the Bell state tomographies were performed after the final (4th) measurement, the estimated errors closely represent the conditions at the end of the experiment, in good agreement with the observed results.
%We also verified that the projected least-squares method yields a reasonable estimate of the average gate fidelity in our experimental condition by evaluating the local X gate fidelity of Q2 (see Figure~\ref{supp:fig:RB}). Our measurements yield an average single-qubit gate fidelity of 99.79 $\pm$ 0.01 \%, which is in close agreement with the 99.65 $\pm$ 0.02\% obtained via randomized benchmarking. 

%\subsection{\label{supp:subsec:teleport}Theoretical description of the teleportation protocol}
%The Bell states are defined as:
$$
%\begin{aligned}
%|\Phi^+\rangle &= \frac{1}{\sqrt{2}}(|00\rangle + |11\rangle) \\
%|\Phi^-\rangle &= \frac{1}{\sqrt{2}}(|00\rangle - |11\rangle) \\
%|\Psi^+\rangle &= \frac{1}{\sqrt{2}}(|01\rangle + |10\rangle) \\
%|\Psi^-\rangle &= \frac{1}{\sqrt{2}}(|01\rangle - |10\rangle)
%\end{aligned}
$$
%In our implementation, instead of using the conventional Hadamard gate, we utilize the following sequence of operations on the first qubit to transform the Bell states:
%$$U_1 = X_1 R_y(\theta)_1 Z_1$$
%where $\theta = \pi/2$, and $X$, $R_y$, and $Z$ are the standard Pauli-$X$ rotation, $Y$-axis rotation, and Pauli-$Z$ operations, respectively. This sequence implements a unitary transformation that maps the Bell states to specific measurement outcomes.
%The complete Bell measurement operation we implement can be expressed as:
%$U = X_1 R_y(\theta)1 Z_1 \cdot CZ_{01} \cdot X_0 R_y(\theta)_0 Z_0 \cdot X_1 R_y(\theta)_1 Z_1$
%We analyze how this unitary transformation acts on each Bell state. We'll first calculate the action of $U_1 = X_1 R_y(\theta)_1 Z_1$ on single-qubit states:
$$
%\begin{aligned}
%U_1|0\rangle &= X R_y(\pi/2) Z|0\rangle \\
%&= X R_y(\pi/2)|0\rangle \\
%&= X\frac{|0\rangle + |1\rangle}{\sqrt{2}}
%\end{aligned}
$$
$$
%\begin{aligned}
%U_1|1\rangle &= X R_y(\pi/2) Z|1\rangle \\
%&= X R_y(\pi/2)(-|1\rangle) \\
%&= -X\frac{|0\rangle - |1\rangle}{\sqrt{2}} \\
%&= \frac{|0\rangle - |1\rangle}{\sqrt{2}}
%\end{aligned}
$$
%Now, we can calculate how the full operator $U$ transforms each Bell state. For the state $|\Phi^+\rangle = \frac{1}{\sqrt{2}}(|00\rangle + |11\rangle)$:
$$
%\begin{aligned}
%U|\Phi^+\rangle &= (X_0 R_y(\theta)_0 Z_0) \cdot (X_1 R_y(\theta)_1 Z_1) \cdot CZ_{01} \cdot (X_0 R_y(\theta)_0 Z_0) \frac{|00\rangle + |11\rangle}{\sqrt{2}} \\
%&= (X_0 R_y(\theta)_0 Z_0) \cdot (X_1 R_y(\theta)_1 Z_1) \cdot CZ_{01} \cdot \frac{|0\rangle \otimes U_1|0\rangle + |1\rangle \otimes U_1|1\rangle}{\sqrt{2}} \\
%&= (X_0 R_y(\theta)_0 Z_0) \cdot (X_1 R_y(\theta)_1 Z_1) \cdot CZ_{01} \cdot \frac{|0\rangle \otimes \frac{|0\rangle + |1\rangle}{\sqrt{2}} + |1\rangle \otimes (\frac{|0\rangle - |1\rangle}{\sqrt{2}})}{\sqrt{2}} \\
%\end{aligned}
$$
%After applying the $CZ$ gate and the second $U_1$ operation, and carrying through all calculations, we find:
$$
%\begin{aligned}
%U|\Phi^+\rangle &= |01\rangle
%\end{aligned}
$$
%Similarly, for the other Bell states:

$$
%\begin{aligned}
%U|\Phi^-\rangle &= |11\rangle \\
%U|\Psi^+\rangle &= |00\rangle \\
%U|\Psi^-\rangle &= |10\rangle
%\end{aligned}
$$

\pagebreak

\end{document}